\documentclass[aps,nofootinbib,prd,superscriptaddress,tightenlines,notitlepage,
twocolumn,showpacs,floatfix]{revtex4-1}
\usepackage[colorlinks]{hyperref}
\usepackage{tabularx}
\usepackage{bm}
\usepackage{graphicx}
\usepackage{amssymb,bm,tensor}
\usepackage{xcolor}
\usepackage{amsmath}
\usepackage[varg]{txfonts}
\usepackage{enumerate}
\usepackage[normalem]{ulem}
\usepackage{bm}
\usepackage[OT1]{fontenc}

\usepackage{graphicx}
\usepackage{dcolumn}
\usepackage{bm}
\usepackage{amsmath}
\usepackage{subfigure}

\let\sss = \scriptscriptstyle
\usepackage[colorlinks]{hyperref}
\usepackage[normalem]{ulem}
\usepackage{xcolor}

\newcommand{\reply}[1]{#1}

\begin{document}

\title{Effects from Dark Matter Halos on X-ray Pulsar Pulse Profiles}

\author{Yukun Liu}
\affiliation{School of Physics, Peking University, Beijing 100871, China}

\author{Hong-Bo Li}
\affiliation{Kavli Institute for Astronomy and Astrophysics, Peking University,
Beijing 100871, China}

\author{Yong Gao}
\affiliation{Max Planck Institute for Gravitational Physics (Albert Einstein Institute), 
Am M\"uhlenberg 1, D-14476 Potsdam-Golm, Germany}

\author{Lijing Shao}\email[Corresponding author: ]{lshao@pku.edu.cn}
\affiliation{Kavli Institute for Astronomy and Astrophysics, Peking University,
Beijing 100871, China}
\affiliation{National Astronomical Observatories, Chinese Academy of Sciences,
Beijing 100012, China}

\author{Zexin Hu}
\affiliation{School of Physics, Peking University, Beijing 100871, China}
\affiliation{Kavli Institute for Astronomy and Astrophysics, Peking University,
Beijing 100871, China}

\date{\today}

\begin{abstract}
Neutron stars (NSs) can capture dark matter (DM) particles because of their deep
gravitational potential and high density. The accumulated DM can affect the
properties of NSs.  In this work we use a general relativistic two-fluid
formalism to solve the structure of DM-admixed NSs (DANSs) and the surrounding
spacetime.  Specifically, we pay attention to the situation where those DANSs
possess DM halos. Due to the gravitational effect of the DM halo, the pulse
profile of an X-ray pulsar is changed. Our study finds a universal relation
between the peak flux deviation of the pulse profile and $M_{\rm halo}/R_{\rm
\sss BM}$, which is the ratio of the DM halo mass, $M_{\rm halo}$, to the 
baryonic matter (BM) core radius, $R_{\rm \sss BM}$. Our results show that, when
$M_{\rm halo}/R_{\rm \sss BM}=0.292$ and the DM particle mass $m_f = 0.3\,$GeV,
the maximum deviation of the profile can be larger than 100$\%$, which has
implication in X-ray pulsar observation.
\end{abstract}

\maketitle

\section{introduction\label{sec:intro}}

Observational evidence has increasingly supported the existence of dark matter
(DM).  For example, DM facilitates the formation of early
galaxies~\cite{Mocz:2019pyf} and helps galaxies rotate at constant speeds in
their outer reaches~\cite{Sofue:2000jx}.  Measurements through gravitational
lensing have revealed some possible DM substructures~\cite{Fadely:2011su}.
According to our current understanding, DM occupies about a quarter of the
energy density of the present Universe, while the baryonic matter (BM)  only
accounts for about 5$\%$~\cite{Planck:2018vyg}.  The nature of DM is a vastly
uncharted territory, and exploring it can help us better understand the
constitution of the Universe.

However, direct methods of searching for DM particles have not yielded
satisfactory results.  Thus it is of great interest to constrain the properties
of DM through  indirect and complementary methods.  Neutron stars (NSs) provide
extreme celestial laboratories for fundamental physics~\cite{Shao:2022koz}. They
may capture DM particles due to their high density and deep gravitational
potential~\cite{Bertone:2007ae, Kouvaris:2010vv, Leung:2011zz, Xiang:2013xwa,
Ellis:2018bkr, Kain:2021hpk, Wang:2021odb, Hippert:2022snq}. Even given a cross
section constraint between DM and BM $\lesssim 10^{-45}\,{\rm cm^2}$ (say, from
underground DM search experiments~\cite{LUX:2016ggv, XENON:2018voc,
PandaX-4T:2021bab, LZ:2022lsv}), the mean free path of DM particles is still
possible to be of the same order as the radius of a NS \cite{Kouvaris:2010vv}.

Different models of DM have different scenarios for DM-admixed NSs
(DANSs)~\cite{Kouvaris:2010vv, Bertone:2007ae, Singh:2022wvw, Liang:2023nvo}.  A
two-fluid model is often used for the equation of state (EOS) of
DANSs~\cite{Comer:1999rs}.  In this model, one fluid is the BM while the other
stands for DM, which can be asymmetric fermionic DM or self-interacting bosonic
DM, and so on.  After solving for the equilibrium equations, a DANS can either
possess a DM core or a DM halo~\cite{Leung:2011zz, Ellis:2018bkr,
Hippert:2022snq, Kain:2021hpk, Xiang:2013xwa}, depending on which fluid has a
larger radius.  DM can significantly affect the properties of NSs. 
\citet{Leung:2011zz} demonstrated that the existence of a fermionic DM core with
the DM particle mass $m_f \sim 1\,$GeV would lower the maximum mass of  NSs.
\citet{Hippert:2022snq} showed that DANSs may have the same mass and radius but
different DM fractions.  \citet{Kain:2021hpk} calculated the \reply{radical} oscillation
frequencies of DANSs and found that the frequencies of DANSs could be larger
than the maximum possible frequencies of normal NSs.
\reply{Axion DM particle stream away from NSs may transfer into photons 
and generate a large broadband contribution to the pulse~\cite{Noordhuis:2022ljw}.
Annihilating DM may affect the thermal conductivity and nuetrino emission 
of NSs~\cite{Cermeno:2017xwb}.}
Numerous investigation
suggests an appealing probability that NSs can be a laboratory for studying DM,
concerning various observables of NSs that can be influenced by DM. To give a
few examples, DM may play a role in binary systems~\cite{Blas:2016ddr,
Malik:2019whk, Pani:2015qhr}; radio telescope may find imprints of DM from NSs'
magnetosphere~\cite{Witte:2021arp, Hook:2018iia, Safdi:2018oeu};  NS surface
temperature can also be impacted by DM~\cite{deLavallaz:2010wp,
Kouvaris:2007ay}. 
More studies on this topic can be found in
Refs.~\cite{Karkevandi:2021ygv, Shakeri:2022dwg, Diedrichs:2023trk,
Deliyergiyev:2023uer, Thakur:2024mxs,Blinnikov:1983gh,Khlopov:1989fj}.

In this work we mainly focus on the effects of DM halos on X-ray pulsar pulse
profiles. A pulsar is a highly magnetized rotating NS that emits electromagnetic
radiations. This radiation can only be observed when the emitting beam points to
the Earth. Hot regions on a pulsar created by inward moving particles colliding
with the NS surface can produce X-ray pulses, which have been observed through
telescopes like the Neutron star Interior Composition ExploreR
(NICER)~\cite{Riley:2019yda, Miller:2019cac, Miller:2021qha,
Vinciguerra:2023qxq}.  The flux of the X-ray pulse changes with time, which
shapes the X-ray pulse profile. It carries information related to the mass and
radius of the NS~\cite{Raaijmakers:2021uju, Hu:2021tyw, Kennedy:2022zml}. If
there is a DM halo around a NS, the pulse profile will be changed due to the
gravitational effects of the massive halo. If such effects are observed, the
properties of DM can be better inferred.

\citet{Miao:2022rqj} and~\citet{shawqi2024interpreting} showed that the X-ray 
profile from a pulsar can be slightly changed if a diffuse DM halo exists.  In
our study, we compare the pulse profile of a DANS with a NS that has the same
total mass, $M$, and the BM core radius, $R_{\rm \sss BM}$, which---for some
pulsars---are inferred observables from orbital dynamics and X-ray thermal
emission, respectively~\cite{Rutledge:2001kc}.  Suppose that the total mass and
the BM radius of a NS are obtained, we can compare its pulse profiles under the
assumptions of a DANS or a normal NS. We find a universal relation between the
peak flux deviation of the DANS pulse profiles and the ratio $M_{\rm
halo}/{R_{\rm \sss BM}}$, where $M_{\rm halo}$ is the mass of the DM halo.
Compared with earlier studies, we consider more comprehensive models,
incorporating a wider range of parameters to provide a thorough analysis of the
impact of DM halos on pulse profiles. We change the position of the hot spots
and the relative direction of observers, showing the fractional deviation of the
peak flux as a function of the colatitude of the emitting region. In addition,
we focus on extended emitting regions besides the point-like approximation,
providing a more realistic representation of the actual situation. In this work,
for the first time we obtain the universal relation that still holds for bosonic
DM model. 

The paper is organized as follows. Section~\ref{sec:formalism} provides an
introduction to the formalism used to determine the structure of DANSs and their
stability.  Section~\ref{sec:DANS} briefly presents some numerical results of
the structure of DANSs.  Section~\ref{sec:profile} outlines general methods for
the calculation of pulse profiles, and shows the influence of fermionic DM halos
in detail. We also extend our study to bosonic DM in this section.
Section~\ref{sec:sum} gives a summary of the paper.

Throughout the paper, we adopt geometric units with $c=G=1$, where $c$ and $G$
denote the speed of light and the gravitational constant respectively. We adopt
$(-\,,+\,,+\,,+)$ for the metric signature. 

\section{Formalism}\label{sec:formalism}

\subsection{The Two-fluid Model and Equation of State}

We briefly review the derivation of the two-fluid \reply{Tolman}–Oppenheimer–Volkoff 
(TOV) equations~\cite{Tolman:1939jz,Oppenheimer:1939ne,Sandin:2008db}.  The
Einstein field equation reads  
\begin{eqnarray}\label{eq:Einstein}
  G^{\mu \nu} = 8\pi T^{\mu\nu} \,,
\end{eqnarray}
where $  G^{\mu \nu} $ and $T^{\mu\nu}$ are the Einstein tensor and
energy-momentum tensor of matters, respectively.  Considering the results of
underground DM search experiments~\cite{LUX:2016ggv, XENON:2018voc,
PandaX-4T:2021bab, LZ:2022lsv}, strict constraints were set for the DM-BM
coupling strength~\cite{Munoz:2003gx}.  For this reason, to the leading order we
neglect all interactions between DM and BM except gravity.  So the
energy-momentum tensor can be divided into the BM component, $T^{\mu\nu}_{\rm
\sss BM}$, and DM component, $T^{\mu\nu}_{\rm \sss DM}$, 
\begin{eqnarray}
  T^{\mu \nu} = T^{\mu \nu}_{\rm \sss BM} + T^{\mu \nu}_{\rm \sss DM} \,.
\end{eqnarray}
Each of them is conserved,
\begin{eqnarray}\label{eq:const}
  T^{\mu \nu}_{ {\rm \sss BM} ;\nu} = T^{\mu \nu}_{ {\rm \sss DM} ;\nu} = 0 \,.
\end{eqnarray}
The two energy-momentum tensors are written as
\begin{eqnarray}
  T^{\mu \nu}_{\rm \sss BM} = (\epsilon _{\rm \sss BM} + p_{\rm \sss
  BM})u^{\mu}_{\rm \sss BM}u^{\nu}_{\rm \sss BM}+p_{\rm \sss BM}g^{\mu\nu} \,,\\
  T^{\mu \nu}_{\rm \sss DM} = (\epsilon _{\rm \sss DM} + p_{\rm \sss
  DM})u^{\mu}_{\rm \sss DM}u^{\nu}_{\rm \sss DM}+p_{\rm \sss DM}g^{\mu\nu} \,.
\end{eqnarray}

Taking the spherical symmetry, we can write the spacetime line element as ${\rm
d} s^2 = -e^{2\mu} {\rm d}t^2 + e^{2\nu} {\rm d} r^2 + r^2({\rm d} \theta^2 +
\sin^2\theta {\rm d}\phi^2)$, where  $\mu$ and $\nu$ are functions of $r$.  From
Eq.~(\ref{eq:Einstein}) and (\ref{eq:const}), the two-fluid TOV equations can be
derived as
\begin{align}\label{eq:TOV}
\frac{{\rm d} m_{i}}{{\rm d} r} &= 4\pi r^2\epsilon _{i} \,,   \\
\frac{{\rm d} p_{i}}{{\rm d} r} &= - \frac{4\pi
r^3p+m}{r^2(1-2m/r)}(\epsilon_{i} + p_{i}) \,,
\end{align}
where the subscripted $i$ denotes the fluid (DM or BM), $p = p_{\rm \sss
DM}+p_{\rm \sss BM}$, and $m = m_{\rm \sss DM}+m_{\rm \sss BM}$.  The particle
number of each component $N_i$ is 
\begin{eqnarray}\label{eq:N}
\frac{{\rm d} N_{i}}{{\rm d} r} = \frac{4\pi r^2n _i}{\sqrt{1-2m/r}} \,,
\end{eqnarray}
where $n_i$ is the particle number density. 

We use the asymmetric fermionic DM model as a prototype, which is often
approximately regarded as a free Fermi gas at zero temperature.  From
thermodynamic statistics, the energy density, pressure, and number density
are~\cite{Shapiro:1983du}
\begin{widetext}
  \begin{eqnarray}
    \epsilon &=& \frac{1}{2\pi^2}\int_{0}^{k_F} k^2\sqrt{k^2+m_f^2} \, {\rm d} k
             = \frac{1}{8\pi^2} \left[ k_F\sqrt{m_f^2+k_F^2} \big(2k_F^2+m_f^2
             \big) - m_f^4{\rm
             ln}\left(\frac{k_F+\sqrt{m_f^2+k_F^2}}{m_f}\right) \right] \,, \\
    p &=& \frac{1}{6\pi^2}\int_{0}^{k_F} \frac{k^4}{\sqrt{k^2+m_f^2}}\, {\rm d}k
    = \frac{1}{24\pi^2}\left[ k_F\sqrt{m_f^2+k_F^2}\big(2k_F^2-3m_f^2\big) + 3m_f^4{\rm
    ln}\left(\frac{k_F+\sqrt{m_f^2+k_F^2}}{m_f}\right) \right] \,, \\ 
    n &=& \frac{k_F^3}{3\pi^2}\,, 
    \label{eq:n}
  \end{eqnarray}  
\end{widetext}
where $k_F$ is the Fermi momentum.  Considering these formulas and eliminating
$k_F$, we obtain $p(\epsilon)$. The EOS of bosonic DM can be found in
Eq.~(\ref{eq:boson}) at the end of Section \ref{sec:profile}. For the EOS of BM,
we use AP4 model for illustration~\cite{Akmal:1997ft}. With this setting, we can
obtain the structure of DANSs and the surrounding spacetime via integration of
the two-fluid TOV equations.

\subsection{The Critical Curve}

An important question is whether a DANS is stable under perturbations.  In the
case of pure NSs, each solution of the TOV equation is uniquely identified by a
single quantity, the central pressure. The entire star becomes unstable if its 
central pressure exceeds the critical point that corresponds to the maximum 
mass~\cite{Shapiro:1983du}. However, when it comes to two-fluid systems, there
are $two$ kinds of fluids and $two$ independent central pressures.  One is the
central pressure of DM, $p_{\rm \sss DM}^{\rm c}$, and the other is the central
pressure of BM, $p_{\rm \sss BM}^{\rm c}$. Thus there is a critical curve
instead of a critical point in the $p_{\rm \sss BM}^{\rm c}$-$p_{\rm \sss
DM}^{\rm c}$ plane. 

The way to get the critical curve is not simply searching for the maximum mass
while fixing one central pressure and varying the other.  The radical
oscillation of a DANS under small perturbations is a problem of the
Sturm-Liouville type, with eigenvalues $\omega_0^2<\omega_1^2<\omega_2^2<\dots$,
where $\omega_0$ is the eigenfrequency of the fundamental mode.  The critical
points with $\omega = 0$ mark the onset of instability~\cite{Shapiro:1983du,
Li:2022qql}.  Because $N_{\rm \sss DM}$ and $N_{\rm \sss BM}$ conserve in
perturbations, \citet{Henriques:1990xg} proposed an equivalent yet more
straightforward criterion,
\begin{eqnarray}
\frac{{\rm d} M}{{\rm d} {\bf p}}
   = \frac{{\rm d} N_{\rm \sss DM}}{{\rm d}{\bf p}} = \frac{{\rm d} N_{\rm \sss
   BM}}{{\rm d} {\bf p}} = 0 \,,
\end{eqnarray}
where $M$ is the total mass, and ${\bf p}$ is a vector that is simultaneously
tangent to both $N_i$ contour lines in the central pressure parameter space. 
\citet{Jetzer:1990xa} has proved that ${\rm d} M$ can be written as a linear 
combination of two ${\rm d} N_i$ at these points. So ${\rm d} M/{\rm d} {\bf p}
= 0$ holds naturally if ${\rm d} N_i/{\rm d}{\bf p} = 0$. 

Using this criterion, we adopt the methods used by \citet{Kain:2021hpk} to
compute the critical curve. This involves computing contour lines of $N_{\rm
\sss DM}$ in the central pressure parameter space using Eq.~(\ref{eq:N}) and
Eq.~(\ref{eq:n}).  Once the contour lines are determined, we follow one line to
search for the point where $M$ reaches its maximum. The critical curve then
consists of all these points.

\section{Structure of dark-matter-admixed neutron stars}\label{sec:DANS}

\begin{figure}[t]
  \includegraphics[scale=0.27]{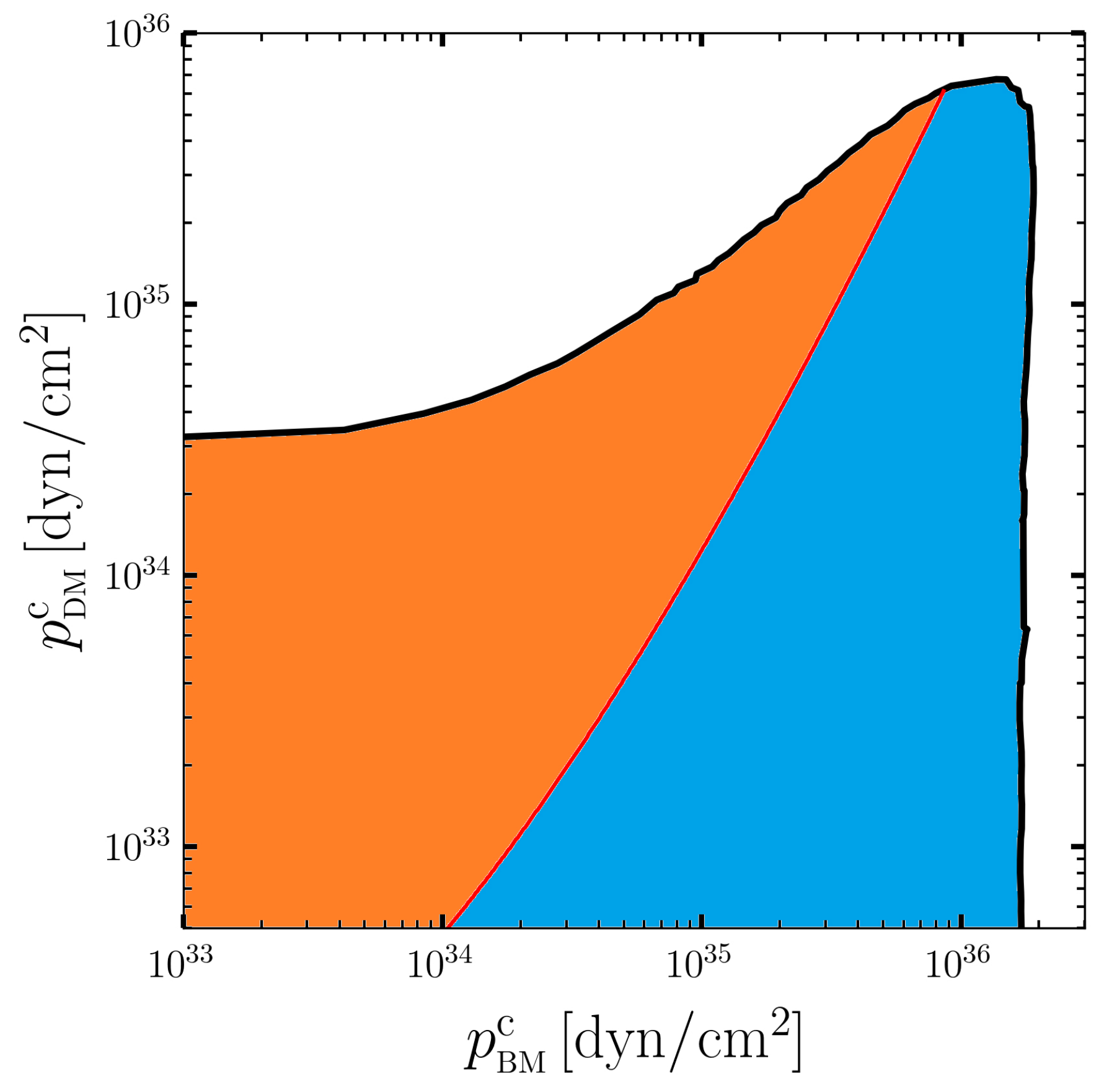}
  \caption{The $p_{\rm \sss BM}^{\rm c}$-$p_{\rm \sss DM}^{\rm c}$ parameter
  space for  solutions in the two-fluid model with $m_f = 0.5\,$GeV.}
  \label{fig:1} 
\end{figure}

By choosing a pair of central pressures as the initial value, we can solve the 
two-fluid TOV equations. The pressures will decrease from the center to the
surfaces at $R_i$, which are defined by $p_i=0$. If $R_{\rm \sss DM} $ is less
than $R_{\rm \sss BM}$, the DANS has a DM core. Conversely, it has a DM halo.
Figure~\ref{fig:1} shows the $p_{\rm \sss BM}^{\rm c}$-$p_{\rm \sss DM}^{\rm c}$
parameter space of DANSs for $m_f = 0.5\,$GeV. The critical curve is drawn with
a thick black line.  The parameter space below the critical curve gives stable
static solutions.  The orange area indicates static solutions with a DM halo,
while the blue area indicates those with a DM core.

\begin{figure}[t]
  \includegraphics[scale=0.45]{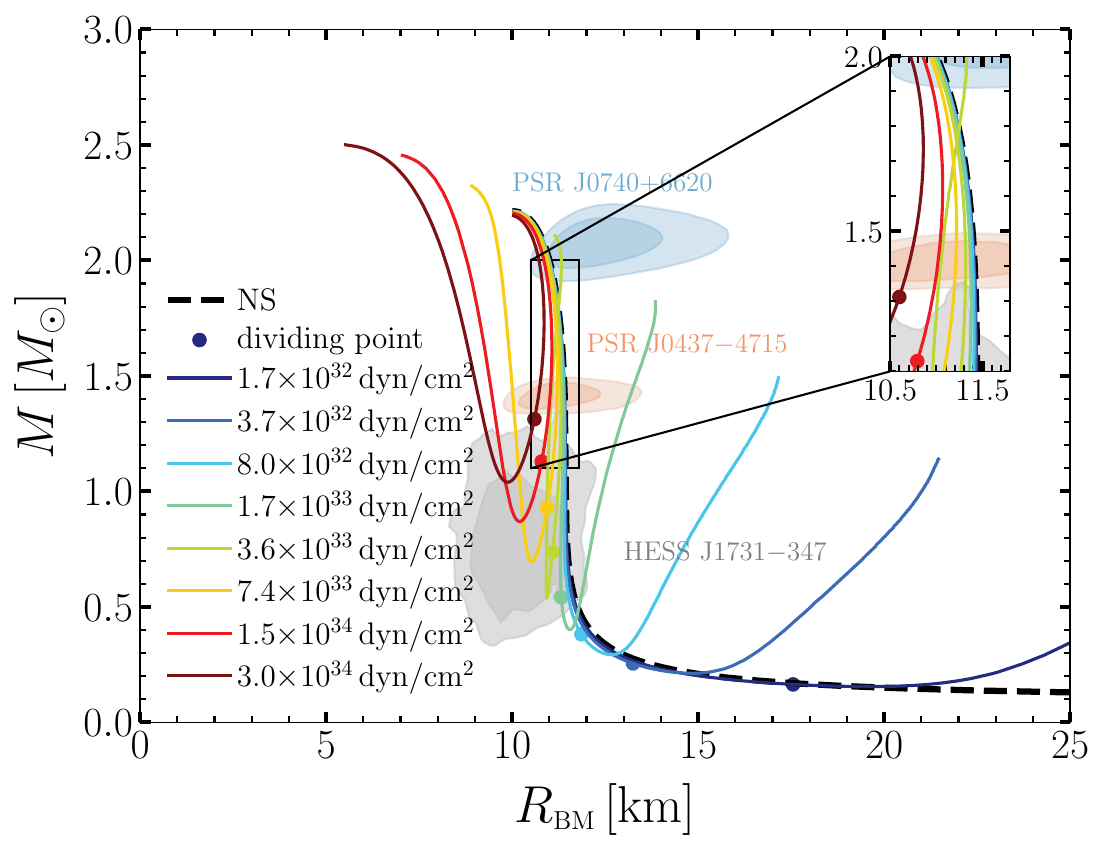}
  \caption{The $M$-$R_{\rm \sss BM}$ relations of DANSs for $m_f = 0.5\,$GeV,
  with different colors indicating different $p_{\rm \sss DM}^{\rm c}$. 
  \reply{NICER constraints are included for PSRs~J0740$+$6620 
  \cite{Riley:2021pdl,Miller:2021qha,Salmi:2024aum} (blue regions), 
  J0437$-$4715 \cite{Amaro-Seoane:2010pks,Choudhury:2024xbk} 
  (orange regions), and HESS~J1731$-$347 \cite{Doroshenko:2022nwp} (grey regions).
  Contours show 68.3$\%$ and 95.4$\%$ 
  credible intervals.} }
  \label{fig:2}
\end{figure}

In the one-fluid system, a 1-dimensional central pressure parameter space
results in a mass-radius curve. Accordingly, a 2-dimensional  $p_{\rm \sss
BM}^{\rm c}$-$p_{\rm \sss DM}^{\rm c}$ parameter space results in a
2-dimensional  $M$-$R_{\rm \sss BM}$ diagram in the two-fluid system.  Here $M$
and $R_{\rm \sss BM}$ are the most essential and straightforward observables.
In Fig.~\ref{fig:2}, we show the relation between $M$ and $R_{\rm \sss BM}$ for
$m_f = 0.5\,$GeV. The thick black dashed line is the mass-radius curve for pure
NSs, and lines with different colors show DANSs. Each line maintains a constant
$p_{\rm \sss DM}^{\rm c}$ while changing $p_{\rm \sss BM}^{\rm c}$.  Meanwhile,
the solid dots are dividing points between solutions with a DM halo and a DM
core. Points closer to the dashed line are solutions with a DM core, while the
other corresponds to solutions with a DM halo.  When $p_{\rm \sss BM}^{\rm c}$ 
gets large enough, a DANS has a DM core, and the lines in the figure get closer 
to the thick black dashed line. That means a DM core affects the structure of
DANSs slightly, just making $M$ and $R_{\rm \sss BM}$ a little smaller compared
to the NSs with the same $p_{\rm \sss BM}^{\rm c}$. When $p_{\rm \sss BM}^{\rm
c}$ is small, a DANS has a DM halo and a significantly different structure. For
a pure NS, a smaller $p_{\rm \sss BM}^{\rm c}$ will result in a larger radius
and a smaller mass. However, the existence of DM will increase the total mass,
and make the BM core shrink further due to the gravity of DM. That is also why
$R_{\rm \sss BM}$ tends to decrease while $M$ tends to increase as $p_{\rm \sss
DM}^{\rm c}$ increases. In addition, different central pressures can result in a
same set of $M$ and $R_{\rm \sss BM}$, as the lines intersecting with each other
in Fig.~\ref{fig:2}.

\begin{figure*}[t]
	\centering
	\subfigure{
		\includegraphics[width=0.45\linewidth]{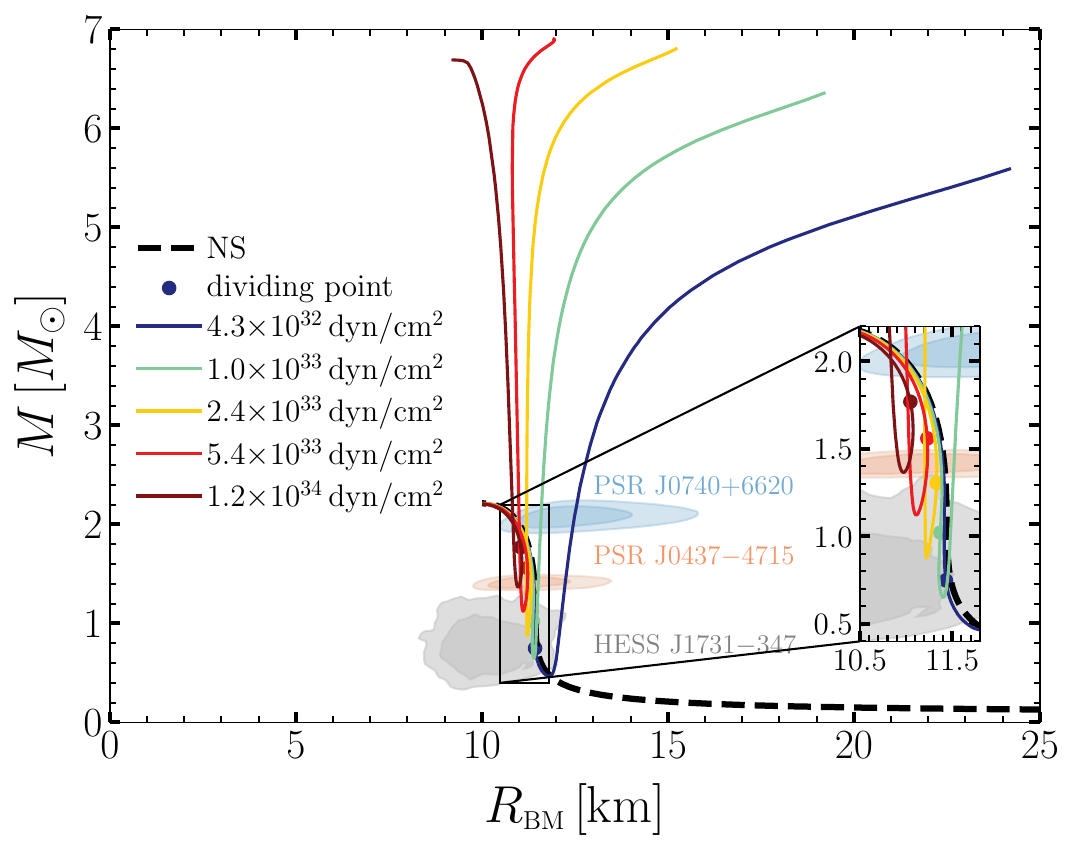}
        \label{subfig:0.3}}
	\subfigure{
		\includegraphics[width=0.45\linewidth,height=0.36\hsize]{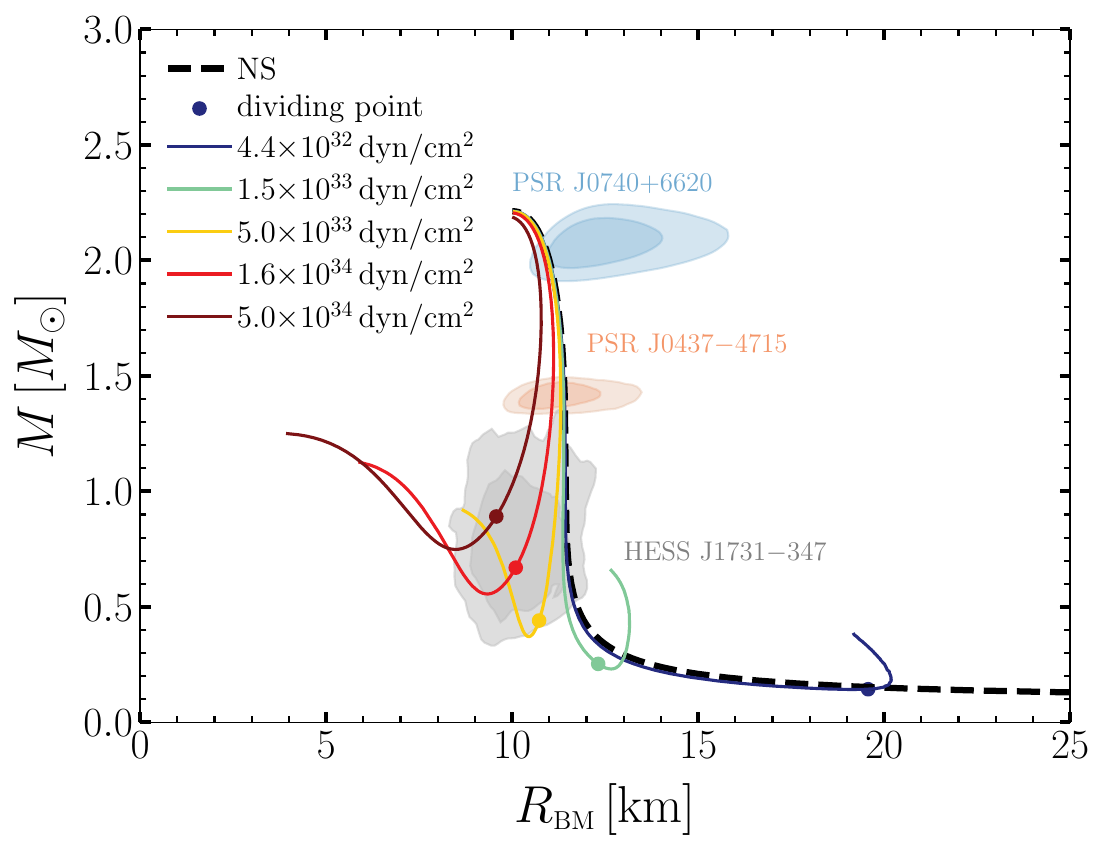}
        \label{subfig:0.7}}
  \caption{The $M$-$R_{\rm \sss BM}$ relations of DANSs.  The left panel is for
  $m_f = 0.3\,$GeV, showing the DM halo region near the dashed line and above.
  The right panel is for $m_f = 0.7\,$GeV, showing the DM halo region compressed
  at the bottom.}
  \label{fig:3}
\end{figure*}

When $m_f = 0.5 \,$GeV, DANSs with either a DM core or a DM halo can be
generated while their masses remain near $1\,M_{\odot}$. Two further examples
with different $m_f$ are given in Fig.~\ref{fig:3}. When $m_f$ is smaller, the
DM halo scenario (solutions with a DM halo) occurs near the dashed line and then
move upwards.  That is because that diffuse DM is easier to form a DM halo and
allow more DM to \reply{exist}.  When $m_f$ is larger, the DM halo scenario will be
compressed at the bottom.

\section{Modification of pulse profiles by dark matter halos}\label{sec:profile}

\subsection{Light Bending Effects of Gravitational Fields}

In this subsection, we briefly outline the methods to calculate the X-ray pulsar
pulse profile.  For detailed derivation, readers can refer
to~\citet{Silva:2018yxz} and~\citet{Xu:2020vbs}.  We assume there is a pair of
hot spots on the BM surface of a DANS with a DM halo.  The radiation from each
of these point-like emitting regions is presumed to be
isotropic~\cite{Miller:1998gr}.  Light emitted from them will be subject to
additional light bending effects due to the gravity of the DM halo. In the
presence of DM, the metric of spacetime outside the BM core is no longer the
Schwarzschild metric, though it is still spherical
symmetric~\cite{Morsink:2007tv}.

\reply{Following Refs.~\cite{Hu:2021tyw,Silva:2018yxz,Xu:2020vbs}, 
we introduce some geometry quantities 
(see Fig.~8 in Ref.~\cite{Xu:2020vbs}) and the calculation of the flux for 
arbitrary spherically symmetric metric.} In the model, a pair of
hot spots is located on the BM surface with colatitudes of $\gamma$ and
$\pi-\gamma$ measured from the spin axis; ${\bm n}$ is a unit vector pointing to
one hot spot from the center of the star; ${\bm k_0}$ is a unit vector in the
direction of light emission; ${\bm k}$ is a unit vector in the direction of
light after bending; ${\bm\beta}$ stands for the velocity of the emitting region
in the local static frame. There are  four other relevant angles: the angle
$\iota$ between the line of sight and the spin axis, the angle $\psi$ between
the line of sight and ${\bm n}$, the angle $\alpha$ between ${\bm k_0}$ and
${\bm n}$, and the angle $\zeta$ between ${\bm k_0}$ and ${\bm
\beta}$~\cite{Xu:2020vbs}.

The observed differential flux in terms of the quantities at the 
emitting region is~\cite{Poutanen:2006hw}
\begin{eqnarray}
  {\rm d}F = \delta ^5 g(R_{\rm \sss BM})I^\prime(\nu^\prime)d\nu^\prime
  \frac{\cos\alpha\sin\alpha} {\sin \psi} \frac{{\rm d} \alpha}{{\rm d}\psi}
  \frac{{\rm d} S^\prime}{D^2} \,,
\end{eqnarray}
where $\delta$ is the Doppler factor, $I^\prime(\nu^\prime)$ is the radiation
intensity as a function of frequency in the comoving frame, ${\rm d} S^\prime$
is the proper differential area at the emitting region, $D$ is the distance
between the pulsar and the observer, and we have defined $-g$ as the
$t\text{-}t$ component of the metric. The Doppler factor, describing how the
boost speed $\beta $ caused by the pulsar rotation affects the flux, can be
written as $ \delta = {\sqrt{1-\beta^2}}/ \big({1-\beta \cos \zeta}
\big)$~\cite{Rybicki:2004hfl}.

Through the analysis in the local static frame~\cite{Poutanen:2006hw}, we can
calculate $\beta$ and $\cos \zeta$ by $ \beta = {R_{\rm \sss BM}\,
\Omega\sin\gamma}/{\sqrt{g(R_{\rm \sss BM})}}$ and $ \cos \zeta = -{\sin \alpha
\sin \iota \sin (\Omega t_{0})}/{\sin \psi}$, where $\Omega$ is the angular
velocity of pulsar rotation, and $t_{0}$ is the coordinate time when the ray is
emitted and has been set to zero when the spot is the closest to the observer.
The relation between $\psi$ and $\alpha$ can be derived by doing an integral
along the path of light propagation~\cite{1983ApJ...274..846P}, $\psi =
\int_{R_{\rm \sss BM}}^{\infty} {\sqrt{fg}} \left[{r^{-2}\sigma^{-2}}-
r^{-4}{g}\right] {\rm {d}}r$, where $f$ is the $r\text{-}r$ component of the
metric and $\sigma \equiv {R_{\rm \sss BM}\sin \alpha}/{\sqrt{g(R_{\rm \sss
BM})}}$ is the impact parameter.

The flux is normalized by a factor 
$\int I^\prime(\nu^\prime)d\nu^\prime {\rm {d}}S^\prime/D^2$, 
\begin{eqnarray}\label{eq:F and g}
  F \equiv \frac{D^2\int dF}{{\rm {d}} S^\prime \int I^\prime(\nu^\prime) {\rm
  {d}} \nu^\prime}
  = \delta^5 g(R_{\rm \sss BM})\frac{\cos\alpha\sin\alpha}{\sin \psi} \frac{{\rm
  {d}}\alpha}{{\rm {d}} \psi}.
\end{eqnarray}
Using geometric analysis, we obtain $\cos \psi = \cos \iota \cos \gamma + \sin
\iota \sin \gamma \sin (\Omega t_0)$.  Thus every emission time $t_0$
corresponds to an angle $\psi$, an impact parameter $\sigma$, and an emission
angle $\alpha$.  Then we calculate the observer time, $t_{\rm obs}$, which is
different from $t_{0}$ because the time of light propagation differs. Taking
this time delay into consideration, we finally
obtained~\cite{1983ApJ...274..846P},
\begin{eqnarray}\label{eq:t}
  t_{\rm obs} = t_{0} + \int_{R_{\rm \sss BM}}^{\infty} \sqrt{\frac{f}{g}}
  \left[\left(1-\frac{\sigma^2g}{r^2}\right)^{-\frac{1}{2}}-1\right] {\rm {d}}r \,.
\end{eqnarray}
Eventually, using Eq.~(\ref{eq:F and g}) and Eq.~(\ref{eq:t}), we obtain the
relation between the normalized flux and the observer time, namely the pulse
profile.

\begin{table*}[htbp]
  \caption{Relevant parameters of two DANSs and corresponding normal NSs, whose 
  $M$ and $R_{\rm \sss BM}$ are the same. The DM particle mass $m_f$ is fixed to
  $0.5\,$GeV. }
  \begin{ruledtabular}\label{tab:1}
    \begin{tabular}{lcccccc}
      Model & $R_{\rm \sss BM}/{\rm km}$ & $R_{\rm \sss DM}/{\rm km}$ &
      $M/M_{\odot}$ & $M_{\rm \sss DM}/M_{\odot}$ & $M_{\rm halo}/M_{\odot}$ &
      $M_{\rm halo}/R_{\rm \sss BM}$\\
      \hline
      DANS1 & 9.5 & 16.9 & 1.2 & 0.44 & 0.11 & 0.016\\
      NS1 & 9.5 & -- & 1.2 & 0 & 0 & 0\\
      \hline
      DANS2 & 11.5 & 51.5 & 1.4 & 1.27 & 1.14 & 0.146\\
      NS2 & 11.5 & -- & 1.4 & 0 & 0 & 0
    \end{tabular}
  \end{ruledtabular}
\end{table*}

\subsection{X-Ray Pulse Profile Flux}

Previously, \citet{Miao:2022rqj} studied the pulse profiles of DANSs, fixing the
BM core radius, $R_{\rm \sss BM}$, and the core mass, $M(R_{\rm \sss BM})$.  To
characterize the effects of DM, they defined a quantity, $M_{\rm halo}/R_{\rm
\sss DM}$, where $M_{\rm halo} = M - M(R_{\rm \sss BM})$, is the mass of the DM
halo. \reply{In fact, it is the half of the increase of the $t\text{-}t$ component of the metric, $g(R_{\rm \sss DM})$, when adding a halo. Here $g$ plays an important role in calculating the flux intensity.}
In our study, as mentioned in Sec.~\ref{sec:intro}, we assume that $M$ and
$R_{\rm \sss BM}$ are potential observables, respectively from orbital dynamics
and X-ray emission modeling.  In this scenario, we find that the flux
deviation---with respect to a normal NS---is better dependent on $M_{\rm
halo}/R_{\rm \sss BM}$ than $M_{\rm halo}/R_{\rm \sss DM}$.  So we choose
$M_{\rm halo}/R_{\rm \sss BM}$ as a characteristic quantity in describing the
effects of DM halos on pulse profiles.  In Table~\ref{tab:1} we choose $m_f =
0.5\,$GeV for illustration, and show two pairs of DANSs (DANS1 and DANS2) and
normal NSs with the same $M$ and $R_{\rm \sss BM}$. DANS1 and DANS2 have similar
$M$ and $R_{\rm \sss BM}$, but DANS1 has a lighter DM halo than DANS2. 

\begin{figure}[t]
  \includegraphics[scale=0.45]{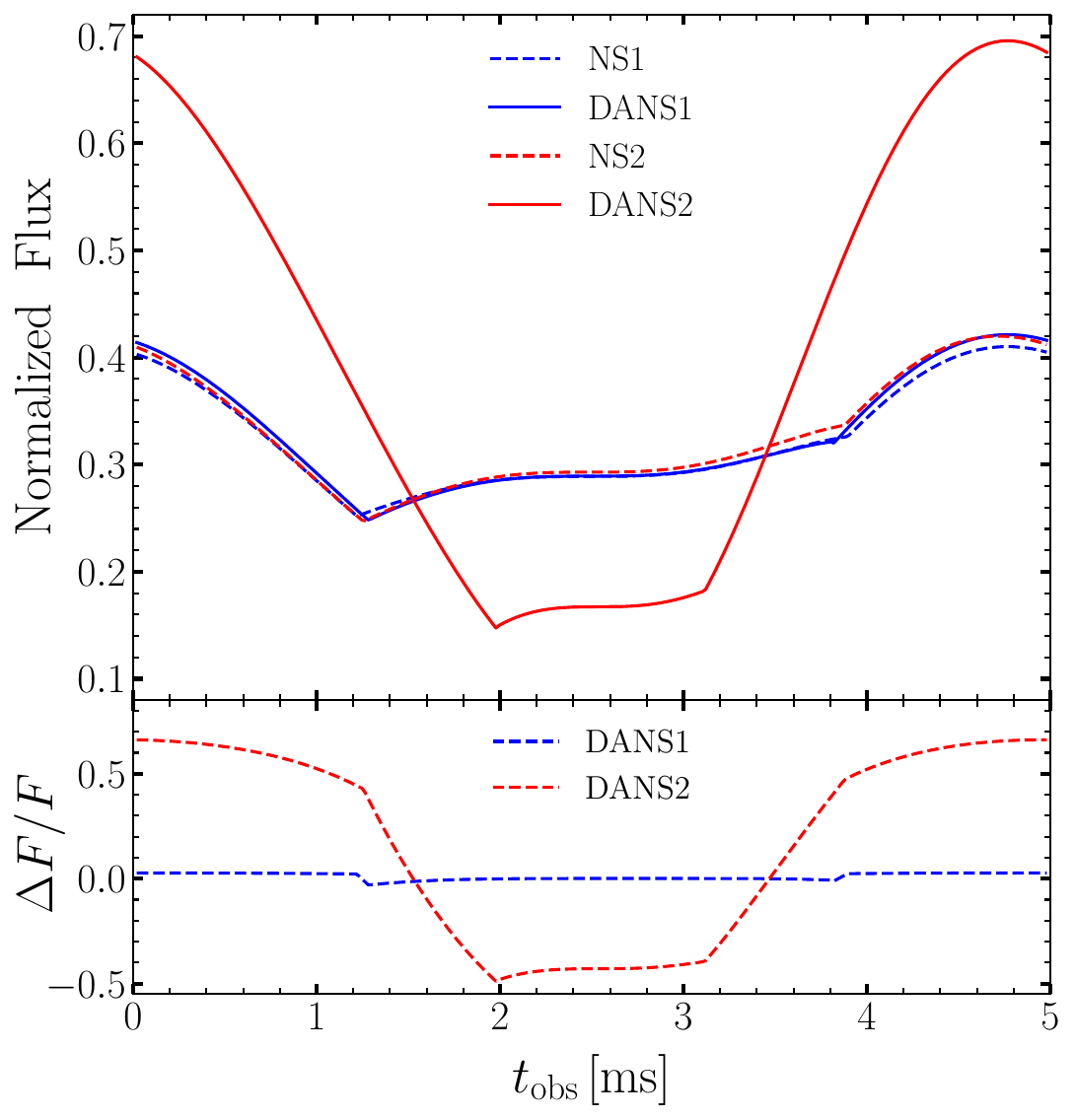}
  \caption{({\it Upper}) Pulse profiles from two point-like hot spots of the
  DANSs and NSs in Table~\ref{tab:1}. The emitting region in the northern
  hemisphere is located at a colatitude $\gamma=45^{\circ}$.  The inclination
  angle of the line of sight is $\iota =45^{\circ}$.  ({\it Lower}) Deviation of
  DANS pulse profiles with respect to the corresponding normal NS pulse
  profiles.}
  \label{fig:4}
\end{figure}

In the upper panel of Fig.~\ref{fig:4}, we show the pulse profiles from 
point-like hot spots with $\gamma = 45^{\circ}$ and $\iota = 45^{\circ}$, from a
pulsar with a rotational period $T = 5\,{\rm ms}$. The normalized flux
deviation, $\Delta F/F \equiv (F_{\rm \sss DANS}-F_{\rm \sss NS})/F_{\rm \sss
NS}$, as a function of $t_{\rm obs}$, is shown in the lower panel of the figure,
where $F_{\rm \sss DANS}$ and $F_{\rm \sss NS}$ are respectively fluxes of DANSs
and normal NSs.  As shown in this figure, the fraction $M_{\rm halo}/R_{\rm \sss
BM}$ significantly affects the DANS pulse profiles. The maximum deviation occurs
near the peak of the pulse profile. Compared to a normal NS, DANS pulse flux
tends to exhibit increased brightness during high-luminosity periods, and
decreased brightness during low-luminosity periods. That can be explained by
Eq.~(\ref{eq:F and g}).  $F$ and $g(R_{\rm \sss BM})$ are positively correlated.
For DANSs, $g(R_{\rm \sss BM})$ is larger because of the existence of DM halos. 
As a result, $F_{\rm \sss DANS}$ will fluctuate in a wider range.  It always
holds true for large $M_{\rm halo}/R_{\rm \sss BM}$ since the difference of
$g(R_{\rm \sss BM})$ between DNASs and NSs is big enough.  However, it is not
shown in this figure that when $M_{\rm halo}/R_{\rm \sss BM}$ drops to about
$10^{-4}$, the amplitude of $F_{\rm \sss DANS}$ may become less than $F_{\rm
\sss NS}$.

\begin{figure}[t]
  \includegraphics[scale=0.45]{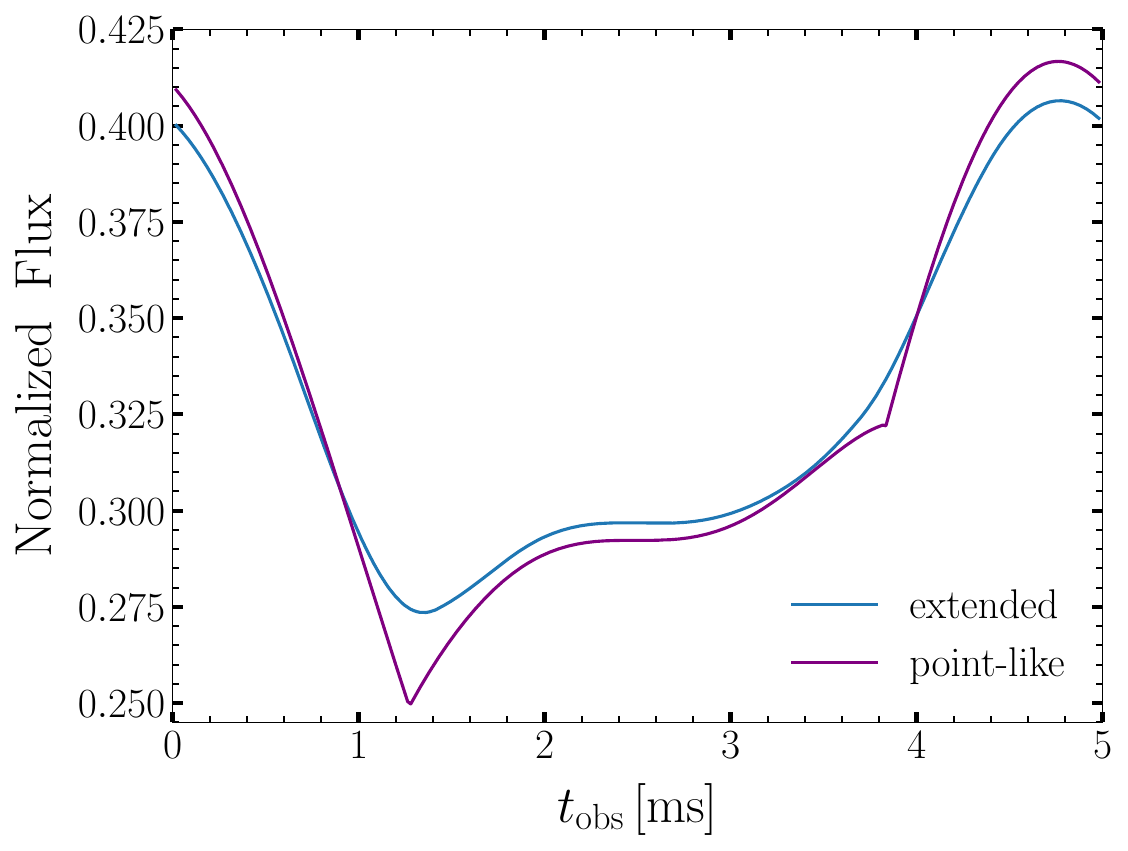}
  \caption{Comparison of pulse profiles of DANS1 with point-like and extended
  emitting regions.}
  \label{fig:6}
\end{figure}

More realistically, we extend our study to a scenario where the emitting regions
are extended. We assume their diameters to be $d = 3\,$km. We consider a grid of
emission spots on the surface~\cite{Bogdanov:2008qm,Hu:2021tyw}.  We use the
same values for $\gamma$ and $\iota$, namely $\gamma=45^{\circ}$ and
$\iota=45^{\circ}$.  To show the differences between point-like emitting regions
and extended emitting regions, we plot a comparison of pulse profiles from DANS1
in Fig.~\ref{fig:6}. An obvious feature of the curve representing the extended
emitting region is the absence of sharp turns. The sharp turn in the point-like
case comes from the emitting region with a colatitude $\gamma = 135^{\circ}$
(the opposite patch).  This region remains invisible when it turns away from the
line of sight.  It only contributes to the overall flux during a part of the
rotational period.  Considering an extended emitting region as an ensemble of
point-like sources, each of them has a distinct  sharp turn.  The addition of
them gives a smooth pulse profile.  

\begin{figure}[t]
  \includegraphics[scale=0.45]{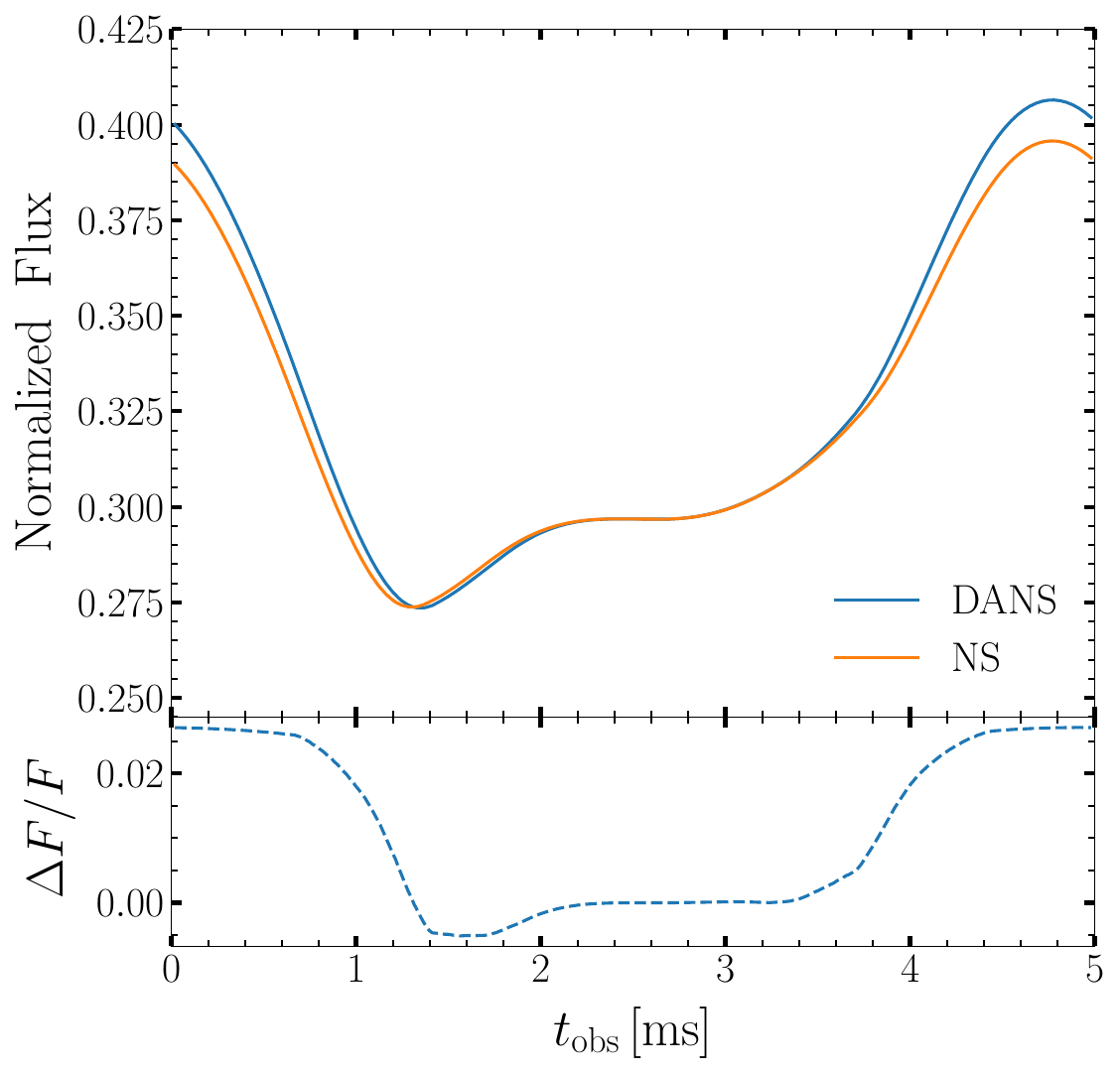}
  \caption{Pulse profiles of DANS1 and NS1 (see Table~\ref{tab:1}) with a pair
  of the emitting region, $\gamma$, and the inclination angle of the line of
  sight, $\iota$.}
  \label{fig:5}
\end{figure}

\begin{figure}[t]
  \includegraphics[scale=0.45]{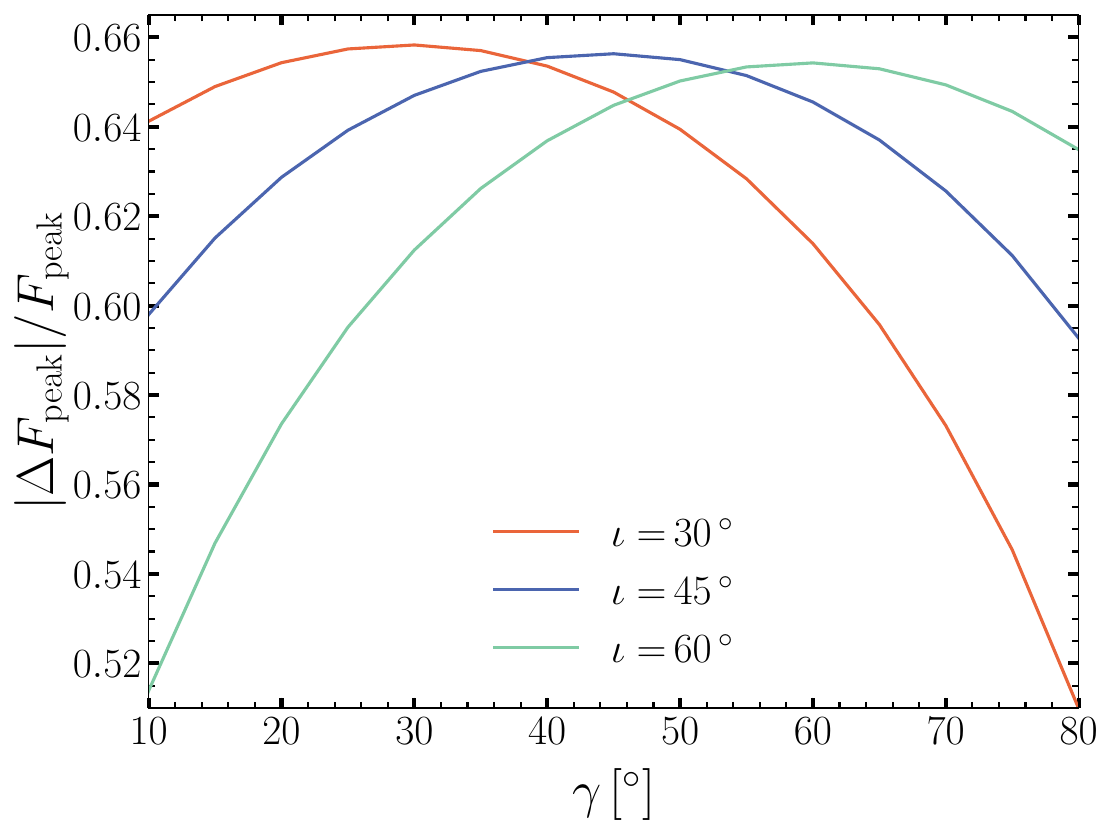}
  \caption{Influence of the colatitude $\gamma$ and the inclination $\iota$ to
  the peak flux deviation of the emitting region between DANS2 and NS2.}
  \label{fig:7}
\end{figure}

To see the effect from DM halos, taking DANS1 and NS1 as example, the pulse
profiles from extended emitting regions are shown in Fig.~\ref{fig:5}.  The
upper panel is their pulse profiles, and the lower panel is $\Delta F/F$.  Our
previously derived conclusions based on point-like emitting regions remain
qualitatively applicable here. Compared with the point-like case, 
the pulse profile from the extended
emitting region exhibits a smaller range between its minimum  and maximum
fluxes. The difference is about 20$\%$ in Fig.~\ref{fig:6}.  To provide an
explanation, we investigate the impact of the colatitude of the emitting region,
$\gamma $, and the observer's relative position, represented by $\iota$.  To see
the deviation more clearly, we take the example of DANS2 and NS2.  The
modification of the peak flux with different $\gamma $ and $\iota$ are shown in
Fig.~\ref{fig:7}.  These two angles exhibit similar impacts on the pulse
profile, as they always appear in pair in the formulas except for the Doppler
factor, which is relatively small. When $\gamma = \iota$, the most significant
deviation of the pulse profile occurs. However, in the context of extended
emitting regions, $\gamma = \iota$ does not hold at each point.  As a result,
the final pulse profile exhibits a gentler overall behavior.

\begin{figure}[t]
  \includegraphics[scale=0.6]{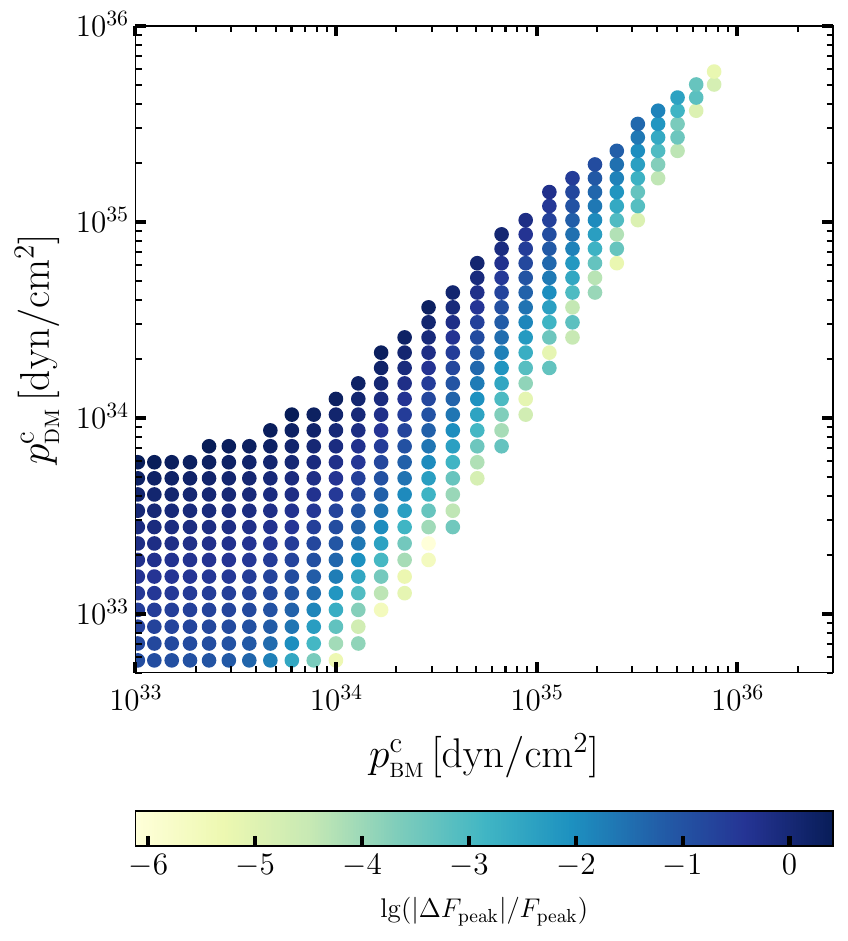}
  \caption{Peak flux deviation in the $p_{\rm \sss BM}^{\rm c}$-$p_{\rm \sss
  DM}^{\rm c}$ parameter space for $m_f = 0.5\,$GeV. Here all DANSs have a DM
  halo. The area with the smallest deviation is near the dividing line between
  the DM halo and the DM core (see Fig.~\ref{fig:1}).} 
  \label{fig:8}
\end{figure}

To provide a comprehensive overview of how DM halos affect pulse profiles, we
compute the peak flux deviations for different central pressures for $m_f =
0.5\,$GeV, assuming the presence of a DM halo.  We only consider DANSs with $
R_{\rm \sss BM} > 3M$, which is a reasonable assumption considering the
\reply{constraints} on the EOS~\cite{1983ApJ...274..846P}.  Our results are depicted in
Fig.~\ref{fig:8}, revealing the substantial influence of DM halos across most
parameter space for DM halos.  When the central pressures lie near the dividing
line  between the DM halo and the DM core (see Fig.~\ref{fig:1}), their pulse
profiles exhibit minimal deviation. Overall, as expected, the peak flux 
deviation tends to increase with an increasing $p_{\rm \sss DM}^{\rm c}$ when 
$p_{\rm \sss BM}^{\rm c}$ is fixed.

\subsection{Universal Relations}

\begin{figure}[t]
  \includegraphics[scale=0.45]{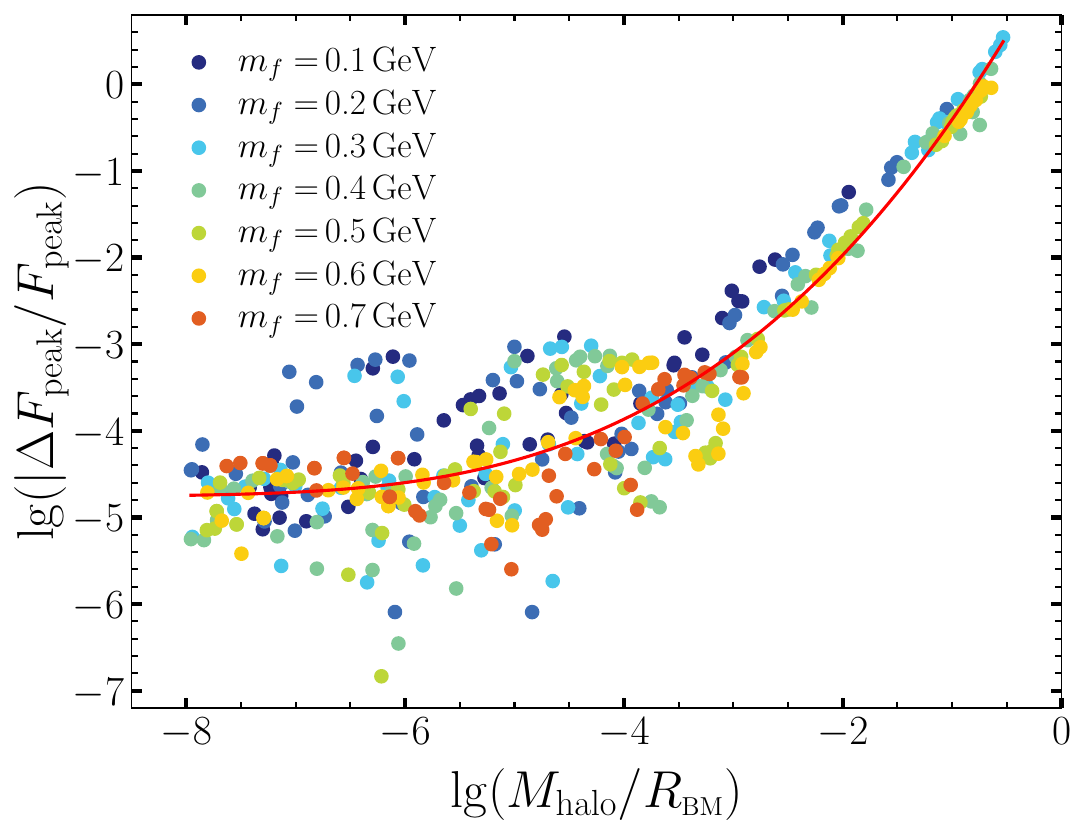}
  \caption{Peak flux deviation as a function of $M_{\rm halo}/R_{\rm \sss BM}$.
  Each point is a randomly chosen DANS. Different colors represent different
  $m_f$.  The solid line is a fitted curve given by Eq.~(\ref{eq:fermion}).} 
  \label{fig:9}
\end{figure}

\reply{The impacts of central pressures on peak flux deviations have been shown in Fig.~\ref{fig:8}. However, we want to derive a more direct relation.}
As mentioned above, the deviation might be highly dependent on the quantity 
$M_{\rm halo}/R_{\rm \sss BM}$. To verify this hypothesis, we calculate a series
of flux deviations systematically with a varying $m_f$. For each $m_f$, we
randomly select $M$ and $R_{\rm \sss BM}$ from the $M$-$R_{\rm \sss BM}$ diagram
where DANSs have halos.  Existing observations have already established
constraints on the mass and radius of NSs~\cite{LIGOScientific:2017vwq,
Bogdanov:2019qjb, Riley:2021pdl, Wolff:2021oba}.  To remain within these bounds,
we only consider cases where $M \in [1,3]\,M_{\odot}$ and $R_{\rm \sss BM} \in
[8,20]\,$km. The colatitude of the northern point-like emitting region $\gamma$
and the inclination angle $\iota$ are randomly distributed in
$[30^{\circ},60^{\circ}]$. 

In Fig.~\ref{fig:9}, we show $|\Delta F_{\rm peak}|/F_{\rm peak}$ against 
$M_{\rm halo}/R_{\rm \sss BM}$ using logarithmic scales. Over a range of $m_f\in
[0.1,0.7]\,$GeV, $M_{\rm halo}/R_{\rm \sss BM}\in [1.1\times10^{-8},
2.9\times10^{-1}]$, all of the randomly selected data points can be loosely
fitted by a cubic function. Defining $f_F(x) = \lg(|\Delta F_{\rm peak}|/F_{\rm
peak})$, $x = \lg(M_{\rm halo}/R_{\rm \sss BM})$, the universal relation is
represented by the following formula,
\begin{eqnarray}\label{eq:fermion}
  f_F(x) = 0.012x^3+0.290x^2+2.351x+1.667 \,.
\end{eqnarray}
In particular, the maximum deviation  of the peak flux exceeds 100$\%$ when 
$M_{\rm halo}/R_{\rm \sss BM} = 0.292$ and $m_f = 0.3\,$GeV, which can
potentially leads to a significant impact on observations. 
\reply{Take PSR~J0740+6620 as an example~\cite{Riley:2021pdl,Miller:2021qha,Salmi:2024aum}. A good constraint on the mass and radius was obtained using about three years’ NICER data, with the total net exposure time about $10^6$\,sec. The uncertainty of the telescope was estimated to be about 10$\%$, that means the flux variations might be observable. }
Analyzing X-ray
pulses of DANSs represents a theoretically viable approach to searching for DM.

\begin{figure}[t]
  \subfigure{\includegraphics[scale=0.46]{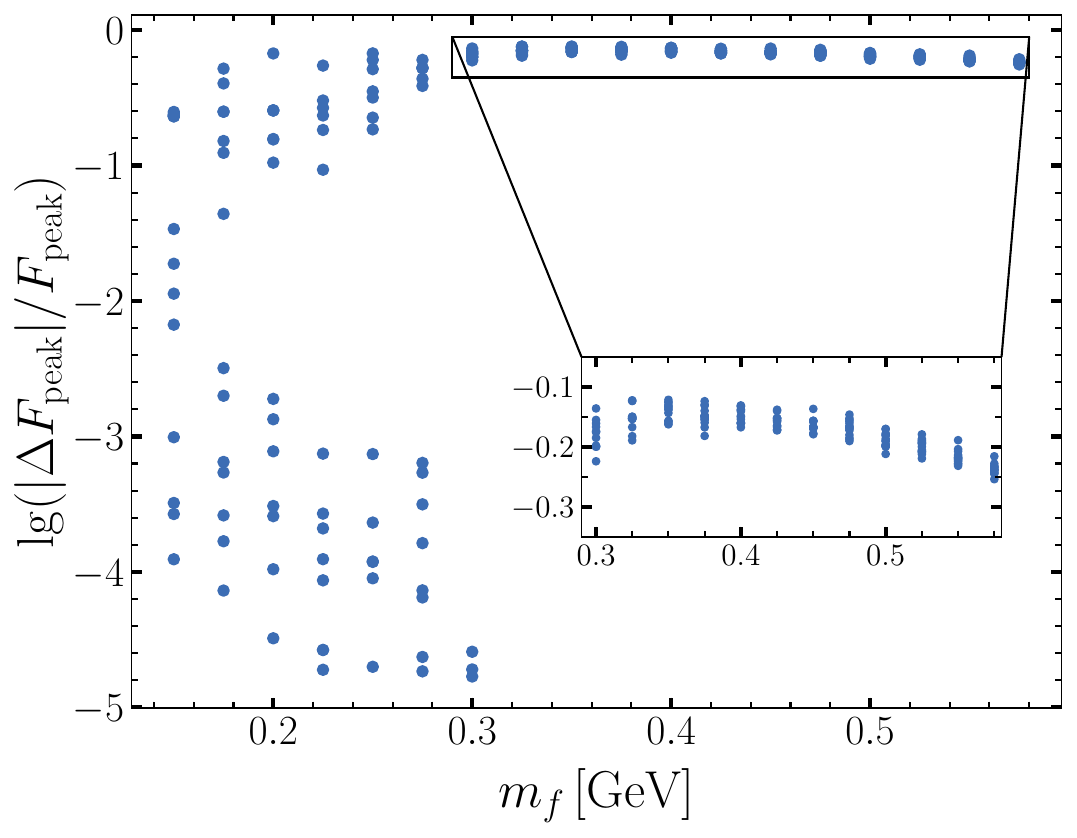}}
  \caption{Peak flux deviation versus the DM particle mass.  There are two
  branches of solutions in the figure: one spans all $m_f$ in the figure, while
  the other stops at $m_f = 0.3\,$GeV.} 
  \label{fig:10}
\end{figure}

We investigate how DM halos affect pulsar pulse profiles of DANSs with typical
NS masses and radii. For a normal NS with a mass of $1.4\,M_{\odot}$, the AP4
EOS yields a radius of $11.4\,$km. We randomly choose DANSs with $M \in
[1.37,1.43]\, M_{\odot}$ and $R_{\rm \sss BM} \in [11.2,11.6]\,$km. The peak
flux deviation versus the DM particle mass $m_f$ is shown in Fig.~\ref{fig:10}.
When $m_f > 0.3\,$GeV, the peak flux deviation remains nearly constant at about
70$\%$. To enhance clarity, we include a smaller inset figure that magnifies
this region.  When $m_f < 0.3\,$GeV, the data points are largely scattered .

It is clear that there are two branches of solutions in Fig.~\ref{fig:10}.  One
of the branches spans all $m_f$ in the figure, while the other stops at $m_f =
0.3\,$GeV. To provide an explanation, we define a ``solution region" where $M
\in [1.37,1.43]\, M_{\odot}$ and $R_{\rm \sss BM} \in [11.2,11.6]\,$km.
Following Figs.~\ref{fig:2}--\ref{fig:3}, for heavy DM particles, say, $m_f =
0.7\,$GeV, the parameter space for DM halo  would not overlap with the solution
region. When $m_f$ is smaller, the data points for DM halos move upwards,
overlapping with the solution region. When $m_f$ continues to decrease, the
dividing points get closer to the black dashed lines (see
Figs.~\ref{fig:2}--\ref{fig:3}), and the second branch of solution occurs.
These solutions are near the dividing points, which means a much more diffuse DM
halo, resulting in a smaller flux deviation.

The closer the solution is to the dividing points, the smaller the flux
deviation is.  When $m_f$ is small, the effect of a diffuse DM halo is
insignificant.  For example, when $M_{\rm halo}/R_{\rm \sss BM} \sim 10^{-2}$ or
$10^{-6}$, according to Fig.~\ref{fig:9}, the quantity $|\Delta F_{\rm
peak}|/F_{\rm peak}$ may differ within $\big[10^{-5}, 10^{-2} \big]$. However,
for a larger $m_f$, there is no DANS with a diffuse DM halo in the solution
region. The quantity $|\Delta F_{\rm peak}|/F_{\rm peak}$ remains in $\big[
10^{-0.3}, 10^{-0.1} \big]$. Consequently, the peak flux deviation induced by a
heavy DM mass can exceed the one induced by a light DM mass by orders of
magnitude, due to significantly different DANS central pressures. These two
varied scenarios manifest as either diffusion of data points for a small $m_f$
or clustering of data points for a large $m_f$ when plotted on a logarithmic
scale.

\begin{figure}[t]
  \includegraphics[scale=0.45]{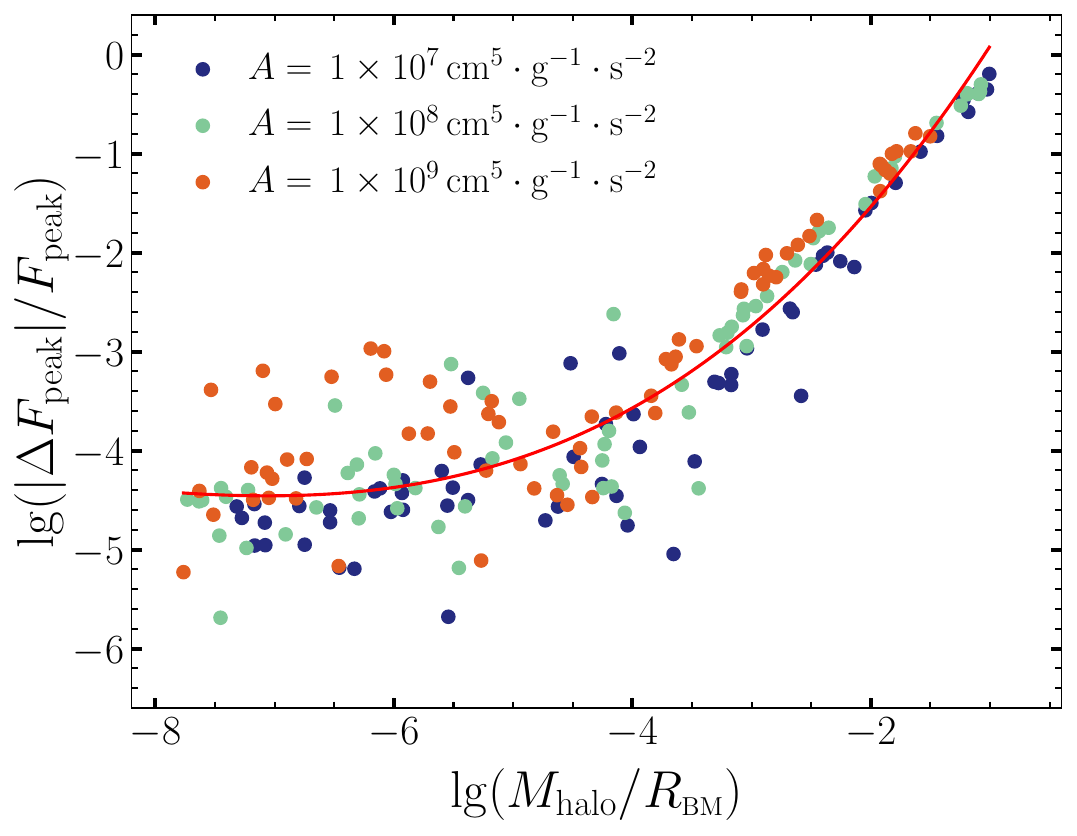}
  \caption{Peak flux deviation as a function of $M_{\rm halo}/R_{\rm \sss BM}$
  for bosonic DM. Different colors are for different $A$ in
  Eq.~(\ref{eq:boson}).  The solid line gives the universal relation in
  Eq.~(\ref{eq:relation}).} 
  \label{fig:11}
\end{figure}

Bosonic DM is another candidate in explaining the DM phenomenology. In a simple
model, the EOS of bosonic DM can be written as~\cite{Li:2012qf, Boehmer:2007um}
\begin{eqnarray}\label{eq:boson}
  p = A\rho^2 \,,
\end{eqnarray}
where $A = 2\pi l/m_f^3$ with $l$ the scattering length.  The universal relation
between $\lg(|\Delta F_{\rm peak}|/F_{\rm peak})$ and $\lg(M_{\rm halo}/R_{\rm
\sss BM})$ for bosonic DM halos is shown in Fig.~\ref{fig:11}. In particular, as
shown by the solid line in the figure, the universal relation can be loosely
represented by the following formula,
\begin{eqnarray}\label{eq:relation}
f_B(x) = 0.010x^3+0.270x^2+2.364x+2.190\,,
\end{eqnarray}
where $f_B(x) = \lg(|\Delta F_{\rm peak}|/F_{\rm peak})$, and $x = \lg(M_{\rm
halo}/R_{\rm \sss BM})$. 

We derive two universal relations concerning fermionic DM and bosonic DM, 
respectively given in Eq.~(\ref{eq:fermion}) and Eq.~(\ref{eq:relation}); also
see Fig.~\ref{fig:9} and Fig.~\ref{fig:11}. Striking similarities can be seen,
which underscores the generality of our results.

\section{conclusions}\label{sec:sum}

In our study, we analyzed the structure of DANSs in a two-fluid model. 
Different $M$-$R_{\rm \sss BM}$ diagrams for DANSs were constructed considering
different DM particle mass, $m_f$. Via integration of the TOV equations, a DANS
possesses either a DM core or a DM halo.  Our primary focus lies in
investigating the impact of DM halos on X-ray pulsar pulse profiles. The pulse
profiles of DANSs were compared with those of normal NSs. In contrast to earlier
studies, we fixed the total mass $M$, instead of the core mass $M(R_{\rm \sss
BM})$, considering that $M$ is a more fundamental and direct observable.
Similarly, we also fixed the BM core radius, $R_{\rm \sss BM}$. 

Our calculations revealed that the presence of dense DM halos will increase the
amplitude of X-ray flux fluctuation. Besides examining point-like hot spots, we
explored the effects of extended emitting regions. We found that extended 
emitting regions tend to reduce the gap between the minimum and maximum fluxes,
by about 20$\%$ (see Fig.~\ref{fig:6}). The impacts of the position of emitting 
regions, represented by the colatitude angle $\gamma$, and the inclination angle
to the line of sight, $\iota$, were also studied.  The influences of $\gamma$
and $\iota$ on the pulse profiles are similar.  When $\gamma = \iota$, the
influence reaches its maximum.

Furthermore, we observed a universal correlation between the peak flux deviation
of DANSs with respect to normal NSs, and the dimensionless quantity, $M_{\rm
halo}/R_{\rm \sss BM}$.  Remarkably, this relation remains consistent when
considering both fermionic and bosonic DM models. \reply{In this paper, we did not consider self-interacting DM, which is a very interesting topic in DM studies (see e.g. Ref.~\cite{Mariani:2023wtv}). 
However, focusing on the universal relation,  \citet{Miao:2022rqj} have studied fermionic self-interacting DM via a 
repulsive Yukawa type, and the relation of theirs holds when the 
interaction strength is randomly chosen. Indepth studies 
concerning different interactions and rich phenomena are left to future investigation.}
We also studied DANSs with
typical masses and radii ($M \simeq 1.4\,M_{\odot}$ and $R_{\rm \sss BM} \simeq
11.4\,{\rm km}$).  It is interesting that the peak flux deviation can be divided
into two different branches, in which the larger deviation stays at about
70$\%$.  In all calculated examples, the maximum deviation of the pulse profiles
can exceed $100\%$ when $M_{\rm halo}/R_{\rm \sss BM}=0.292$ and $m_f =
0.3\,$GeV, leading to a potentially significant observable difference. Our study
suggests that X-ray observation of pulsars may be a feasible way to search for
DM. 

\begin{acknowledgments}
This work was supported by the Beijing Natural Science Foundation (QY23009,
1242018), the National SKA Program of China (2020SKA0120300), the National
Natural Science Foundation of China (11991053),  the Max Planck Partner Group
Program funded by the Max Planck Society, and the High-Performance Computing
Platform of Peking University.
\end{acknowledgments}


\bibliography{pulse}

\begin{thebibliography}{75}%
\makeatletter
\providecommand \@ifxundefined [1]{%
 \@ifx{#1\undefined}
}%
\providecommand \@ifnum [1]{%
 \ifnum #1\expandafter \@firstoftwo
 \else \expandafter \@secondoftwo
 \fi
}%
\providecommand \@ifx [1]{%
 \ifx #1\expandafter \@firstoftwo
 \else \expandafter \@secondoftwo
 \fi
}%
\providecommand \natexlab [1]{#1}%
\providecommand \enquote  [1]{``#1''}%
\providecommand \bibnamefont  [1]{#1}%
\providecommand \bibfnamefont [1]{#1}%
\providecommand \citenamefont [1]{#1}%
\providecommand \href@noop [0]{\@secondoftwo}%
\providecommand \href [0]{\begingroup \@sanitize@url \@href}%
\providecommand \@href[1]{\@@startlink{#1}\@@href}%
\providecommand \@@href[1]{\endgroup#1\@@endlink}%
\providecommand \@sanitize@url [0]{\catcode `\\12\catcode `\$12\catcode
  `\&12\catcode `\#12\catcode `\^12\catcode `\_12\catcode `\%12\relax}%
\providecommand \@@startlink[1]{}%
\providecommand \@@endlink[0]{}%
\providecommand \url  [0]{\begingroup\@sanitize@url \@url }%
\providecommand \@url [1]{\endgroup\@href {#1}{\urlprefix }}%
\providecommand \urlprefix  [0]{URL }%
\providecommand \Eprint [0]{\href }%
\providecommand \doibase [0]{http://dx.doi.org/}%
\providecommand \selectlanguage [0]{\@gobble}%
\providecommand \bibinfo  [0]{\@secondoftwo}%
\providecommand \bibfield  [0]{\@secondoftwo}%
\providecommand \translation [1]{[#1]}%
\providecommand \BibitemOpen [0]{}%
\providecommand \bibitemStop [0]{}%
\providecommand \bibitemNoStop [0]{.\EOS\space}%
\providecommand \EOS [0]{\spacefactor3000\relax}%
\providecommand \BibitemShut  [1]{\csname bibitem#1\endcsname}%
\let\auto@bib@innerbib\@empty
\bibitem [{\citenamefont {Mocz}\ \emph {et~al.}(2019)\citenamefont {Mocz} \emph
  {et~al.}}]{Mocz:2019pyf}%
  \BibitemOpen
  \bibfield  {author} {\bibinfo {author} {\bibfnamefont {P.}~\bibnamefont
  {Mocz}} \emph {et~al.},\ }\href {\doibase 10.1103/PhysRevLett.123.141301}
  {\bibfield  {journal} {\bibinfo  {journal} {Phys. Rev. Lett.}\ }\textbf
  {\bibinfo {volume} {123}},\ \bibinfo {pages} {141301} (\bibinfo {year}
  {2019})},\ \Eprint {http://arxiv.org/abs/1910.01653} {arXiv:1910.01653
  [astro-ph.GA]} \BibitemShut {NoStop}%
\bibitem [{\citenamefont {Sofue}\ and\ \citenamefont
  {Rubin}(2001)}]{Sofue:2000jx}%
  \BibitemOpen
  \bibfield  {author} {\bibinfo {author} {\bibfnamefont {Y.}~\bibnamefont
  {Sofue}}\ and\ \bibinfo {author} {\bibfnamefont {V.}~\bibnamefont {Rubin}},\
  }\href {\doibase 10.1146/annurev.astro.39.1.137} {\bibfield  {journal}
  {\bibinfo  {journal} {Ann. Rev. Astron. Astrophys.}\ }\textbf {\bibinfo
  {volume} {39}},\ \bibinfo {pages} {137} (\bibinfo {year} {2001})},\ \Eprint
  {http://arxiv.org/abs/astro-ph/0010594} {arXiv:astro-ph/0010594} \BibitemShut
  {NoStop}%
\bibitem [{\citenamefont {Fadely}\ and\ \citenamefont
  {Keeton}(2012)}]{Fadely:2011su}%
  \BibitemOpen
  \bibfield  {author} {\bibinfo {author} {\bibfnamefont {R.}~\bibnamefont
  {Fadely}}\ and\ \bibinfo {author} {\bibfnamefont {C.~R.}\ \bibnamefont
  {Keeton}},\ }\href {\doibase 10.1111/j.1365-2966.2011.19729.x} {\bibfield
  {journal} {\bibinfo  {journal} {Mon. Not. Roy. Astron. Soc.}\ }\textbf
  {\bibinfo {volume} {419}},\ \bibinfo {pages} {936} (\bibinfo {year}
  {2012})},\ \Eprint {http://arxiv.org/abs/1109.0548} {arXiv:1109.0548
  [astro-ph.CO]} \BibitemShut {NoStop}%
\bibitem [{\citenamefont {Aghanim}\ \emph {et~al.}(2020)\citenamefont {Aghanim}
  \emph {et~al.}}]{Planck:2018vyg}%
  \BibitemOpen
  \bibfield  {author} {\bibinfo {author} {\bibfnamefont {N.}~\bibnamefont
  {Aghanim}} \emph {et~al.} (\bibinfo {collaboration} {Planck}),\ }\href
  {\doibase 10.1051/0004-6361/201833910} {\bibfield  {journal} {\bibinfo
  {journal} {Astron. Astrophys.}\ }\textbf {\bibinfo {volume} {641}},\ \bibinfo
  {pages} {A6} (\bibinfo {year} {2020})},\ \bibinfo {note} {[Erratum:
  Astron.Astrophys. 652, C4 (2021)]},\ \Eprint
  {http://arxiv.org/abs/1807.06209} {arXiv:1807.06209 [astro-ph.CO]}
  \BibitemShut {NoStop}%
\bibitem [{\citenamefont {Shao}\ and\ \citenamefont
  {Yagi}(2022)}]{Shao:2022koz}%
  \BibitemOpen
  \bibfield  {author} {\bibinfo {author} {\bibfnamefont {L.}~\bibnamefont
  {Shao}}\ and\ \bibinfo {author} {\bibfnamefont {K.}~\bibnamefont {Yagi}},\
  }\href {\doibase 10.1016/j.scib.2022.09.018} {\bibfield  {journal} {\bibinfo
  {journal} {Sci. Bull.}\ }\textbf {\bibinfo {volume} {67}},\ \bibinfo {pages}
  {1946} (\bibinfo {year} {2022})},\ \Eprint {http://arxiv.org/abs/2209.03351}
  {arXiv:2209.03351 [gr-qc]} \BibitemShut {NoStop}%
\bibitem [{\citenamefont {Bertone}\ and\ \citenamefont
  {Fairbairn}(2008)}]{Bertone:2007ae}%
  \BibitemOpen
  \bibfield  {author} {\bibinfo {author} {\bibfnamefont {G.}~\bibnamefont
  {Bertone}}\ and\ \bibinfo {author} {\bibfnamefont {M.}~\bibnamefont
  {Fairbairn}},\ }\href {\doibase 10.1103/PhysRevD.77.043515} {\bibfield
  {journal} {\bibinfo  {journal} {Phys. Rev. D}\ }\textbf {\bibinfo {volume}
  {77}},\ \bibinfo {pages} {043515} (\bibinfo {year} {2008})},\ \Eprint
  {http://arxiv.org/abs/0709.1485} {arXiv:0709.1485 [astro-ph]} \BibitemShut
  {NoStop}%
\bibitem [{\citenamefont {Kouvaris}\ and\ \citenamefont
  {Tinyakov}(2010)}]{Kouvaris:2010vv}%
  \BibitemOpen
  \bibfield  {author} {\bibinfo {author} {\bibfnamefont {C.}~\bibnamefont
  {Kouvaris}}\ and\ \bibinfo {author} {\bibfnamefont {P.}~\bibnamefont
  {Tinyakov}},\ }\href {\doibase 10.1103/PhysRevD.82.063531} {\bibfield
  {journal} {\bibinfo  {journal} {Phys. Rev. D}\ }\textbf {\bibinfo {volume}
  {82}},\ \bibinfo {pages} {063531} (\bibinfo {year} {2010})},\ \Eprint
  {http://arxiv.org/abs/1004.0586} {arXiv:1004.0586 [astro-ph.GA]} \BibitemShut
  {NoStop}%
\bibitem [{\citenamefont {Leung}\ \emph {et~al.}(2011)\citenamefont {Leung},
  \citenamefont {Chu},\ and\ \citenamefont {Lin}}]{Leung:2011zz}%
  \BibitemOpen
  \bibfield  {author} {\bibinfo {author} {\bibfnamefont {S.~C.}\ \bibnamefont
  {Leung}}, \bibinfo {author} {\bibfnamefont {M.~C.}\ \bibnamefont {Chu}}, \
  and\ \bibinfo {author} {\bibfnamefont {L.~M.}\ \bibnamefont {Lin}},\ }\href
  {\doibase 10.1103/PhysRevD.84.107301} {\bibfield  {journal} {\bibinfo
  {journal} {Phys. Rev. D}\ }\textbf {\bibinfo {volume} {84}},\ \bibinfo
  {pages} {107301} (\bibinfo {year} {2011})},\ \Eprint
  {http://arxiv.org/abs/1111.1787} {arXiv:1111.1787 [astro-ph.CO]} \BibitemShut
  {NoStop}%
\bibitem [{\citenamefont {Xiang}\ \emph {et~al.}(2014)\citenamefont {Xiang},
  \citenamefont {Jiang}, \citenamefont {Zhang},\ and\ \citenamefont
  {Yang}}]{Xiang:2013xwa}%
  \BibitemOpen
  \bibfield  {author} {\bibinfo {author} {\bibfnamefont {Q.-F.}\ \bibnamefont
  {Xiang}}, \bibinfo {author} {\bibfnamefont {W.-Z.}\ \bibnamefont {Jiang}},
  \bibinfo {author} {\bibfnamefont {D.-R.}\ \bibnamefont {Zhang}}, \ and\
  \bibinfo {author} {\bibfnamefont {R.-Y.}\ \bibnamefont {Yang}},\ }\href
  {\doibase 10.1103/PhysRevC.89.025803} {\bibfield  {journal} {\bibinfo
  {journal} {Phys. Rev. C}\ }\textbf {\bibinfo {volume} {89}},\ \bibinfo
  {pages} {025803} (\bibinfo {year} {2014})},\ \Eprint
  {http://arxiv.org/abs/1305.7354} {arXiv:1305.7354 [astro-ph.SR]} \BibitemShut
  {NoStop}%
\bibitem [{\citenamefont {Ellis}\ \emph {et~al.}(2018)\citenamefont {Ellis},
  \citenamefont {H\"utsi}, \citenamefont {Kannike}, \citenamefont {Marzola},
  \citenamefont {Raidal},\ and\ \citenamefont {Vaskonen}}]{Ellis:2018bkr}%
  \BibitemOpen
  \bibfield  {author} {\bibinfo {author} {\bibfnamefont {J.}~\bibnamefont
  {Ellis}}, \bibinfo {author} {\bibfnamefont {G.}~\bibnamefont {H\"utsi}},
  \bibinfo {author} {\bibfnamefont {K.}~\bibnamefont {Kannike}}, \bibinfo
  {author} {\bibfnamefont {L.}~\bibnamefont {Marzola}}, \bibinfo {author}
  {\bibfnamefont {M.}~\bibnamefont {Raidal}}, \ and\ \bibinfo {author}
  {\bibfnamefont {V.}~\bibnamefont {Vaskonen}},\ }\href {\doibase
  10.1103/PhysRevD.97.123007} {\bibfield  {journal} {\bibinfo  {journal} {Phys.
  Rev. D}\ }\textbf {\bibinfo {volume} {97}},\ \bibinfo {pages} {123007}
  (\bibinfo {year} {2018})},\ \Eprint {http://arxiv.org/abs/1804.01418}
  {arXiv:1804.01418 [astro-ph.CO]} \BibitemShut {NoStop}%
\bibitem [{\citenamefont {Kain}(2021)}]{Kain:2021hpk}%
  \BibitemOpen
  \bibfield  {author} {\bibinfo {author} {\bibfnamefont {B.}~\bibnamefont
  {Kain}},\ }\href {\doibase 10.1103/PhysRevD.103.043009} {\bibfield  {journal}
  {\bibinfo  {journal} {Phys. Rev. D}\ }\textbf {\bibinfo {volume} {103}},\
  \bibinfo {pages} {043009} (\bibinfo {year} {2021})},\ \Eprint
  {http://arxiv.org/abs/2102.08257} {arXiv:2102.08257 [gr-qc]} \BibitemShut
  {NoStop}%
\bibitem [{\citenamefont {Wang}\ \emph {et~al.}(2021)\citenamefont {Wang},
  \citenamefont {Miao},\ and\ \citenamefont {Shao}}]{Wang:2021odb}%
  \BibitemOpen
  \bibfield  {author} {\bibinfo {author} {\bibfnamefont {Y.}~\bibnamefont
  {Wang}}, \bibinfo {author} {\bibfnamefont {X.}~\bibnamefont {Miao}}, \ and\
  \bibinfo {author} {\bibfnamefont {L.}~\bibnamefont {Shao}},\ }\href {\doibase
  10.15940/j.cnki.0001-5245.2021.05.008} {\bibfield  {journal} {\bibinfo
  {journal} {Acta Astron. Sin.}\ }\textbf {\bibinfo {volume} {62}},\ \bibinfo
  {pages} {97} (\bibinfo {year} {2021})}\BibitemShut {NoStop}%
\bibitem [{\citenamefont {Hippert}\ \emph {et~al.}(2023)\citenamefont
  {Hippert}, \citenamefont {Dillingham}, \citenamefont {Tan}, \citenamefont
  {Curtin}, \citenamefont {Noronha-Hostler},\ and\ \citenamefont
  {Yunes}}]{Hippert:2022snq}%
  \BibitemOpen
  \bibfield  {author} {\bibinfo {author} {\bibfnamefont {M.}~\bibnamefont
  {Hippert}}, \bibinfo {author} {\bibfnamefont {E.}~\bibnamefont {Dillingham}},
  \bibinfo {author} {\bibfnamefont {H.}~\bibnamefont {Tan}}, \bibinfo {author}
  {\bibfnamefont {D.}~\bibnamefont {Curtin}}, \bibinfo {author} {\bibfnamefont
  {J.}~\bibnamefont {Noronha-Hostler}}, \ and\ \bibinfo {author} {\bibfnamefont
  {N.}~\bibnamefont {Yunes}},\ }\href {\doibase 10.1103/PhysRevD.107.115028}
  {\bibfield  {journal} {\bibinfo  {journal} {Phys. Rev. D}\ }\textbf {\bibinfo
  {volume} {107}},\ \bibinfo {pages} {115028} (\bibinfo {year} {2023})},\
  \Eprint {http://arxiv.org/abs/2211.08590} {arXiv:2211.08590 [astro-ph.HE]}
  \BibitemShut {NoStop}%
\bibitem [{\citenamefont {Akerib}\ \emph {et~al.}(2017)\citenamefont {Akerib}
  \emph {et~al.}}]{LUX:2016ggv}%
  \BibitemOpen
  \bibfield  {author} {\bibinfo {author} {\bibfnamefont {D.~S.}\ \bibnamefont
  {Akerib}} \emph {et~al.} (\bibinfo {collaboration} {LUX}),\ }\href {\doibase
  10.1103/PhysRevLett.118.021303} {\bibfield  {journal} {\bibinfo  {journal}
  {Phys. Rev. Lett.}\ }\textbf {\bibinfo {volume} {118}},\ \bibinfo {pages}
  {021303} (\bibinfo {year} {2017})},\ \Eprint
  {http://arxiv.org/abs/1608.07648} {arXiv:1608.07648 [astro-ph.CO]}
  \BibitemShut {NoStop}%
\bibitem [{\citenamefont {Aprile}\ \emph {et~al.}(2018)\citenamefont {Aprile}
  \emph {et~al.}}]{XENON:2018voc}%
  \BibitemOpen
  \bibfield  {author} {\bibinfo {author} {\bibfnamefont {E.}~\bibnamefont
  {Aprile}} \emph {et~al.} (\bibinfo {collaboration} {XENON}),\ }\href
  {\doibase 10.1103/PhysRevLett.121.111302} {\bibfield  {journal} {\bibinfo
  {journal} {Phys. Rev. Lett.}\ }\textbf {\bibinfo {volume} {121}},\ \bibinfo
  {pages} {111302} (\bibinfo {year} {2018})},\ \Eprint
  {http://arxiv.org/abs/1805.12562} {arXiv:1805.12562 [astro-ph.CO]}
  \BibitemShut {NoStop}%
\bibitem [{\citenamefont {Meng}\ \emph {et~al.}(2021)\citenamefont {Meng} \emph
  {et~al.}}]{PandaX-4T:2021bab}%
  \BibitemOpen
  \bibfield  {author} {\bibinfo {author} {\bibfnamefont {Y.}~\bibnamefont
  {Meng}} \emph {et~al.} (\bibinfo {collaboration} {PandaX-4T}),\ }\href
  {\doibase 10.1103/PhysRevLett.127.261802} {\bibfield  {journal} {\bibinfo
  {journal} {Phys. Rev. Lett.}\ }\textbf {\bibinfo {volume} {127}},\ \bibinfo
  {pages} {261802} (\bibinfo {year} {2021})},\ \Eprint
  {http://arxiv.org/abs/2107.13438} {arXiv:2107.13438 [hep-ex]} \BibitemShut
  {NoStop}%
\bibitem [{\citenamefont {Aalbers}\ \emph {et~al.}(2023)\citenamefont {Aalbers}
  \emph {et~al.}}]{LZ:2022lsv}%
  \BibitemOpen
  \bibfield  {author} {\bibinfo {author} {\bibfnamefont {J.}~\bibnamefont
  {Aalbers}} \emph {et~al.} (\bibinfo {collaboration} {LZ}),\ }\href {\doibase
  10.1103/PhysRevLett.131.041002} {\bibfield  {journal} {\bibinfo  {journal}
  {Phys. Rev. Lett.}\ }\textbf {\bibinfo {volume} {131}},\ \bibinfo {pages}
  {041002} (\bibinfo {year} {2023})},\ \Eprint
  {http://arxiv.org/abs/2207.03764} {arXiv:2207.03764 [hep-ex]} \BibitemShut
  {NoStop}%
\bibitem [{\citenamefont {Singh}\ \emph {et~al.}(2023)\citenamefont {Singh},
  \citenamefont {Gupta}, \citenamefont {Berti}, \citenamefont {Reddy},\ and\
  \citenamefont {Sathyaprakash}}]{Singh:2022wvw}%
  \BibitemOpen
  \bibfield  {author} {\bibinfo {author} {\bibfnamefont {D.}~\bibnamefont
  {Singh}}, \bibinfo {author} {\bibfnamefont {A.}~\bibnamefont {Gupta}},
  \bibinfo {author} {\bibfnamefont {E.}~\bibnamefont {Berti}}, \bibinfo
  {author} {\bibfnamefont {S.}~\bibnamefont {Reddy}}, \ and\ \bibinfo {author}
  {\bibfnamefont {B.~S.}\ \bibnamefont {Sathyaprakash}},\ }\href {\doibase
  10.1103/PhysRevD.107.083037} {\bibfield  {journal} {\bibinfo  {journal}
  {Phys. Rev. D}\ }\textbf {\bibinfo {volume} {107}},\ \bibinfo {pages}
  {083037} (\bibinfo {year} {2023})},\ \Eprint
  {http://arxiv.org/abs/2210.15739} {arXiv:2210.15739 [gr-qc]} \BibitemShut
  {NoStop}%
\bibitem [{\citenamefont {Liang}\ and\ \citenamefont
  {Shao}(2023)}]{Liang:2023nvo}%
  \BibitemOpen
  \bibfield  {author} {\bibinfo {author} {\bibfnamefont {D.}~\bibnamefont
  {Liang}}\ and\ \bibinfo {author} {\bibfnamefont {L.}~\bibnamefont {Shao}},\
  }\href {\doibase 10.1088/1475-7516/2023/08/016} {\bibfield  {journal}
  {\bibinfo  {journal} {JCAP}\ }\textbf {\bibinfo {volume} {08}},\ \bibinfo
  {pages} {016} (\bibinfo {year} {2023})},\ \Eprint
  {http://arxiv.org/abs/2303.05107} {arXiv:2303.05107 [astro-ph.HE]}
  \BibitemShut {NoStop}%
\bibitem [{\citenamefont {Comer}\ \emph {et~al.}(1999)\citenamefont {Comer},
  \citenamefont {Langlois},\ and\ \citenamefont {Lin}}]{Comer:1999rs}%
  \BibitemOpen
  \bibfield  {author} {\bibinfo {author} {\bibfnamefont {G.~L.}\ \bibnamefont
  {Comer}}, \bibinfo {author} {\bibfnamefont {D.}~\bibnamefont {Langlois}}, \
  and\ \bibinfo {author} {\bibfnamefont {L.~M.}\ \bibnamefont {Lin}},\ }\href
  {\doibase 10.1103/PhysRevD.60.104025} {\bibfield  {journal} {\bibinfo
  {journal} {Phys. Rev. D}\ }\textbf {\bibinfo {volume} {60}},\ \bibinfo
  {pages} {104025} (\bibinfo {year} {1999})},\ \Eprint
  {http://arxiv.org/abs/gr-qc/9908040} {arXiv:gr-qc/9908040} \BibitemShut
  {NoStop}%
\bibitem [{\citenamefont {Noordhuis}\ \emph {et~al.}(2023)\citenamefont
  {Noordhuis}, \citenamefont {Prabhu}, \citenamefont {Witte}, \citenamefont
  {Chen}, \citenamefont {Cruz},\ and\ \citenamefont
  {Weniger}}]{Noordhuis:2022ljw}%
  \BibitemOpen
  \bibfield  {author} {\bibinfo {author} {\bibfnamefont {D.}~\bibnamefont
  {Noordhuis}}, \bibinfo {author} {\bibfnamefont {A.}~\bibnamefont {Prabhu}},
  \bibinfo {author} {\bibfnamefont {S.~J.}\ \bibnamefont {Witte}}, \bibinfo
  {author} {\bibfnamefont {A.~Y.}\ \bibnamefont {Chen}}, \bibinfo {author}
  {\bibfnamefont {F.}~\bibnamefont {Cruz}}, \ and\ \bibinfo {author}
  {\bibfnamefont {C.}~\bibnamefont {Weniger}},\ }\href {\doibase
  10.1103/PhysRevLett.131.111004} {\bibfield  {journal} {\bibinfo  {journal}
  {Phys. Rev. Lett.}\ }\textbf {\bibinfo {volume} {131}},\ \bibinfo {pages}
  {111004} (\bibinfo {year} {2023})},\ \Eprint
  {http://arxiv.org/abs/2209.09917} {arXiv:2209.09917 [hep-ph]} \BibitemShut
  {NoStop}%
\bibitem [{\citenamefont {Cerme\~no}\ \emph {et~al.}(2017)\citenamefont
  {Cerme\~no}, \citenamefont {P\'erez-Garc\'\i{}a},\ and\ \citenamefont
  {Silk}}]{Cermeno:2017xwb}%
  \BibitemOpen
  \bibfield  {author} {\bibinfo {author} {\bibfnamefont {M.}~\bibnamefont
  {Cerme\~no}}, \bibinfo {author} {\bibfnamefont {M.~A.}\ \bibnamefont
  {P\'erez-Garc\'\i{}a}}, \ and\ \bibinfo {author} {\bibfnamefont
  {J.}~\bibnamefont {Silk}},\ }\href {\doibase 10.1017/pasa.2017.38} {\bibfield
   {journal} {\bibinfo  {journal} {Publ. Astron. Soc. Austral.}\ }\textbf
  {\bibinfo {volume} {34}},\ \bibinfo {pages} {e043} (\bibinfo {year}
  {2017})},\ \Eprint {http://arxiv.org/abs/1710.06866} {arXiv:1710.06866
  [astro-ph.HE]} \BibitemShut {NoStop}%
\bibitem [{\citenamefont {Blas}\ \emph {et~al.}(2017)\citenamefont {Blas},
  \citenamefont {Nacir},\ and\ \citenamefont {Sibiryakov}}]{Blas:2016ddr}%
  \BibitemOpen
  \bibfield  {author} {\bibinfo {author} {\bibfnamefont {D.}~\bibnamefont
  {Blas}}, \bibinfo {author} {\bibfnamefont {D.~L.}\ \bibnamefont {Nacir}}, \
  and\ \bibinfo {author} {\bibfnamefont {S.}~\bibnamefont {Sibiryakov}},\
  }\href {\doibase 10.1103/PhysRevLett.118.261102} {\bibfield  {journal}
  {\bibinfo  {journal} {Phys. Rev. Lett.}\ }\textbf {\bibinfo {volume} {118}},\
  \bibinfo {pages} {261102} (\bibinfo {year} {2017})},\ \Eprint
  {http://arxiv.org/abs/1612.06789} {arXiv:1612.06789 [hep-ph]} \BibitemShut
  {NoStop}%
\bibitem [{\citenamefont {Malik}\ \emph {et~al.}(2019)\citenamefont {Malik},
  \citenamefont {Agrawal}, \citenamefont {De}, \citenamefont {Samaddar},
  \citenamefont {Provid\^encia}, \citenamefont {Mondal},\ and\ \citenamefont
  {Jha}}]{Malik:2019whk}%
  \BibitemOpen
  \bibfield  {author} {\bibinfo {author} {\bibfnamefont {T.}~\bibnamefont
  {Malik}}, \bibinfo {author} {\bibfnamefont {B.~K.}\ \bibnamefont {Agrawal}},
  \bibinfo {author} {\bibfnamefont {J.~N.}\ \bibnamefont {De}}, \bibinfo
  {author} {\bibfnamefont {S.~K.}\ \bibnamefont {Samaddar}}, \bibinfo {author}
  {\bibfnamefont {C.}~\bibnamefont {Provid\^encia}}, \bibinfo {author}
  {\bibfnamefont {C.}~\bibnamefont {Mondal}}, \ and\ \bibinfo {author}
  {\bibfnamefont {T.~K.}\ \bibnamefont {Jha}},\ }\href {\doibase
  10.1103/PhysRevC.99.052801} {\bibfield  {journal} {\bibinfo  {journal} {Phys.
  Rev. C}\ }\textbf {\bibinfo {volume} {99}},\ \bibinfo {pages} {052801}
  (\bibinfo {year} {2019})},\ \Eprint {http://arxiv.org/abs/1901.04371}
  {arXiv:1901.04371 [nucl-th]} \BibitemShut {NoStop}%
\bibitem [{\citenamefont {Pani}(2015)}]{Pani:2015qhr}%
  \BibitemOpen
  \bibfield  {author} {\bibinfo {author} {\bibfnamefont {P.}~\bibnamefont
  {Pani}},\ }\href {\doibase 10.1103/PhysRevD.92.123530} {\bibfield  {journal}
  {\bibinfo  {journal} {Phys. Rev. D}\ }\textbf {\bibinfo {volume} {92}},\
  \bibinfo {pages} {123530} (\bibinfo {year} {2015})},\ \Eprint
  {http://arxiv.org/abs/1512.01236} {arXiv:1512.01236 [astro-ph.HE]}
  \BibitemShut {NoStop}%
\bibitem [{\citenamefont {Witte}\ \emph {et~al.}(2021)\citenamefont {Witte},
  \citenamefont {Noordhuis}, \citenamefont {Edwards},\ and\ \citenamefont
  {Weniger}}]{Witte:2021arp}%
  \BibitemOpen
  \bibfield  {author} {\bibinfo {author} {\bibfnamefont {S.~J.}\ \bibnamefont
  {Witte}}, \bibinfo {author} {\bibfnamefont {D.}~\bibnamefont {Noordhuis}},
  \bibinfo {author} {\bibfnamefont {T.~D.~P.}\ \bibnamefont {Edwards}}, \ and\
  \bibinfo {author} {\bibfnamefont {C.}~\bibnamefont {Weniger}},\ }\href
  {\doibase 10.1103/PhysRevD.104.103030} {\bibfield  {journal} {\bibinfo
  {journal} {Phys. Rev. D}\ }\textbf {\bibinfo {volume} {104}},\ \bibinfo
  {pages} {103030} (\bibinfo {year} {2021})},\ \Eprint
  {http://arxiv.org/abs/2104.07670} {arXiv:2104.07670 [hep-ph]} \BibitemShut
  {NoStop}%
\bibitem [{\citenamefont {Hook}\ \emph {et~al.}(2018)\citenamefont {Hook},
  \citenamefont {Kahn}, \citenamefont {Safdi},\ and\ \citenamefont
  {Sun}}]{Hook:2018iia}%
  \BibitemOpen
  \bibfield  {author} {\bibinfo {author} {\bibfnamefont {A.}~\bibnamefont
  {Hook}}, \bibinfo {author} {\bibfnamefont {Y.}~\bibnamefont {Kahn}}, \bibinfo
  {author} {\bibfnamefont {B.~R.}\ \bibnamefont {Safdi}}, \ and\ \bibinfo
  {author} {\bibfnamefont {Z.}~\bibnamefont {Sun}},\ }\href {\doibase
  10.1103/PhysRevLett.121.241102} {\bibfield  {journal} {\bibinfo  {journal}
  {Phys. Rev. Lett.}\ }\textbf {\bibinfo {volume} {121}},\ \bibinfo {pages}
  {241102} (\bibinfo {year} {2018})},\ \Eprint
  {http://arxiv.org/abs/1804.03145} {arXiv:1804.03145 [hep-ph]} \BibitemShut
  {NoStop}%
\bibitem [{\citenamefont {Safdi}\ \emph {et~al.}(2019)\citenamefont {Safdi},
  \citenamefont {Sun},\ and\ \citenamefont {Chen}}]{Safdi:2018oeu}%
  \BibitemOpen
  \bibfield  {author} {\bibinfo {author} {\bibfnamefont {B.~R.}\ \bibnamefont
  {Safdi}}, \bibinfo {author} {\bibfnamefont {Z.}~\bibnamefont {Sun}}, \ and\
  \bibinfo {author} {\bibfnamefont {A.~Y.}\ \bibnamefont {Chen}},\ }\href
  {\doibase 10.1103/PhysRevD.99.123021} {\bibfield  {journal} {\bibinfo
  {journal} {Phys. Rev. D}\ }\textbf {\bibinfo {volume} {99}},\ \bibinfo
  {pages} {123021} (\bibinfo {year} {2019})},\ \Eprint
  {http://arxiv.org/abs/1811.01020} {arXiv:1811.01020 [astro-ph.CO]}
  \BibitemShut {NoStop}%
\bibitem [{\citenamefont {de~Lavallaz}\ and\ \citenamefont
  {Fairbairn}(2010)}]{deLavallaz:2010wp}%
  \BibitemOpen
  \bibfield  {author} {\bibinfo {author} {\bibfnamefont {A.}~\bibnamefont
  {de~Lavallaz}}\ and\ \bibinfo {author} {\bibfnamefont {M.}~\bibnamefont
  {Fairbairn}},\ }\href {\doibase 10.1103/PhysRevD.81.123521} {\bibfield
  {journal} {\bibinfo  {journal} {Phys. Rev. D}\ }\textbf {\bibinfo {volume}
  {81}},\ \bibinfo {pages} {123521} (\bibinfo {year} {2010})},\ \Eprint
  {http://arxiv.org/abs/1004.0629} {arXiv:1004.0629 [astro-ph.GA]} \BibitemShut
  {NoStop}%
\bibitem [{\citenamefont {Kouvaris}(2008)}]{Kouvaris:2007ay}%
  \BibitemOpen
  \bibfield  {author} {\bibinfo {author} {\bibfnamefont {C.}~\bibnamefont
  {Kouvaris}},\ }\href {\doibase 10.1103/PhysRevD.77.023006} {\bibfield
  {journal} {\bibinfo  {journal} {Phys. Rev. D}\ }\textbf {\bibinfo {volume}
  {77}},\ \bibinfo {pages} {023006} (\bibinfo {year} {2008})},\ \Eprint
  {http://arxiv.org/abs/0708.2362} {arXiv:0708.2362 [astro-ph]} \BibitemShut
  {NoStop}%
\bibitem [{\citenamefont {Karkevandi}\ \emph {et~al.}(2022)\citenamefont
  {Karkevandi}, \citenamefont {Shakeri}, \citenamefont {Sagun},\ and\
  \citenamefont {Ivanytskyi}}]{Karkevandi:2021ygv}%
  \BibitemOpen
  \bibfield  {author} {\bibinfo {author} {\bibfnamefont {D.~R.}\ \bibnamefont
  {Karkevandi}}, \bibinfo {author} {\bibfnamefont {S.}~\bibnamefont {Shakeri}},
  \bibinfo {author} {\bibfnamefont {V.}~\bibnamefont {Sagun}}, \ and\ \bibinfo
  {author} {\bibfnamefont {O.}~\bibnamefont {Ivanytskyi}},\ }\href {\doibase
  10.1103/PhysRevD.105.023001} {\bibfield  {journal} {\bibinfo  {journal}
  {Phys. Rev. D}\ }\textbf {\bibinfo {volume} {105}},\ \bibinfo {pages}
  {023001} (\bibinfo {year} {2022})},\ \Eprint
  {http://arxiv.org/abs/2109.03801} {arXiv:2109.03801 [astro-ph.HE]}
  \BibitemShut {NoStop}%
\bibitem [{\citenamefont {Shakeri}\ and\ \citenamefont
  {Karkevandi}(2024)}]{Shakeri:2022dwg}%
  \BibitemOpen
  \bibfield  {author} {\bibinfo {author} {\bibfnamefont {S.}~\bibnamefont
  {Shakeri}}\ and\ \bibinfo {author} {\bibfnamefont {D.~R.}\ \bibnamefont
  {Karkevandi}},\ }\href {\doibase 10.1103/PhysRevD.109.043029} {\bibfield
  {journal} {\bibinfo  {journal} {Phys. Rev. D}\ }\textbf {\bibinfo {volume}
  {109}},\ \bibinfo {pages} {043029} (\bibinfo {year} {2024})},\ \Eprint
  {http://arxiv.org/abs/2210.17308} {arXiv:2210.17308 [astro-ph.HE]}
  \BibitemShut {NoStop}%
\bibitem [{\citenamefont {Diedrichs}\ \emph {et~al.}(2023)\citenamefont
  {Diedrichs}, \citenamefont {Becker}, \citenamefont {Jockel}, \citenamefont
  {Christian}, \citenamefont {Sagunski},\ and\ \citenamefont
  {Schaffner-Bielich}}]{Diedrichs:2023trk}%
  \BibitemOpen
  \bibfield  {author} {\bibinfo {author} {\bibfnamefont {R.~F.}\ \bibnamefont
  {Diedrichs}}, \bibinfo {author} {\bibfnamefont {N.}~\bibnamefont {Becker}},
  \bibinfo {author} {\bibfnamefont {C.}~\bibnamefont {Jockel}}, \bibinfo
  {author} {\bibfnamefont {J.-E.}\ \bibnamefont {Christian}}, \bibinfo {author}
  {\bibfnamefont {L.}~\bibnamefont {Sagunski}}, \ and\ \bibinfo {author}
  {\bibfnamefont {J.}~\bibnamefont {Schaffner-Bielich}},\ }\href {\doibase
  10.1103/PhysRevD.108.064009} {\bibfield  {journal} {\bibinfo  {journal}
  {Phys. Rev. D}\ }\textbf {\bibinfo {volume} {108}},\ \bibinfo {pages}
  {064009} (\bibinfo {year} {2023})},\ \Eprint
  {http://arxiv.org/abs/2303.04089} {arXiv:2303.04089 [gr-qc]} \BibitemShut
  {NoStop}%
\bibitem [{\citenamefont {Deliyergiyev}\ \emph {et~al.}(2023)\citenamefont
  {Deliyergiyev}, \citenamefont {Del~Popolo},\ and\ \citenamefont
  {Delliou}}]{Deliyergiyev:2023uer}%
  \BibitemOpen
  \bibfield  {author} {\bibinfo {author} {\bibfnamefont {M.}~\bibnamefont
  {Deliyergiyev}}, \bibinfo {author} {\bibfnamefont {A.}~\bibnamefont
  {Del~Popolo}}, \ and\ \bibinfo {author} {\bibfnamefont {M.~L.}\ \bibnamefont
  {Delliou}},\ }\href {\doibase 10.1093/mnras/stad3311} {\bibfield  {journal}
  {\bibinfo  {journal} {Mon. Not. Roy. Astron. Soc.}\ }\textbf {\bibinfo
  {volume} {527}},\ \bibinfo {pages} {4483} (\bibinfo {year} {2023})},\ \Eprint
  {http://arxiv.org/abs/2311.00113} {arXiv:2311.00113 [astro-ph.GA]}
  \BibitemShut {NoStop}%
\bibitem [{\citenamefont {Thakur}\ \emph {et~al.}(2024)\citenamefont {Thakur},
  \citenamefont {Malik},\ and\ \citenamefont {Jha}}]{Thakur:2024mxs}%
  \BibitemOpen
  \bibfield  {author} {\bibinfo {author} {\bibfnamefont {P.}~\bibnamefont
  {Thakur}}, \bibinfo {author} {\bibfnamefont {T.}~\bibnamefont {Malik}}, \
  and\ \bibinfo {author} {\bibfnamefont {T.~K.}\ \bibnamefont {Jha}},\ }\href
  {\doibase 10.3390/particles7010005} {\bibfield  {journal} {\bibinfo
  {journal} {Particles}\ }\textbf {\bibinfo {volume} {7}},\ \bibinfo {pages}
  {80} (\bibinfo {year} {2024})},\ \Eprint {http://arxiv.org/abs/2401.07773}
  {arXiv:2401.07773 [hep-ph]} \BibitemShut {NoStop}%
\bibitem [{\citenamefont {Blinnikov}\ and\ \citenamefont
  {Khlopov}(1983)}]{Blinnikov:1983gh}%
  \BibitemOpen
  \bibfield  {author} {\bibinfo {author} {\bibfnamefont {S.~I.}\ \bibnamefont
  {Blinnikov}}\ and\ \bibinfo {author} {\bibfnamefont {M.}~\bibnamefont
  {Khlopov}},\ }\href@noop {} {\bibfield  {journal} {\bibinfo  {journal} {Sov.
  Astron.}\ }\textbf {\bibinfo {volume} {27}},\ \bibinfo {pages} {371}
  (\bibinfo {year} {1983})}\BibitemShut {NoStop}%
\bibitem [{\citenamefont {Khlopov}\ \emph {et~al.}(1991)\citenamefont
  {Khlopov}, \citenamefont {Beskin}, \citenamefont {Bochkarev}, \citenamefont
  {Pustylnik},\ and\ \citenamefont {Pustylnik}}]{Khlopov:1989fj}%
  \BibitemOpen
  \bibfield  {author} {\bibinfo {author} {\bibfnamefont {M.~Y.}\ \bibnamefont
  {Khlopov}}, \bibinfo {author} {\bibfnamefont {G.~M.}\ \bibnamefont {Beskin}},
  \bibinfo {author} {\bibfnamefont {N.~E.}\ \bibnamefont {Bochkarev}}, \bibinfo
  {author} {\bibfnamefont {L.~A.}\ \bibnamefont {Pustylnik}}, \ and\ \bibinfo
  {author} {\bibfnamefont {S.~A.}\ \bibnamefont {Pustylnik}},\ }\href@noop {}
  {\bibfield  {journal} {\bibinfo  {journal} {Sov. Astron.}\ }\textbf {\bibinfo
  {volume} {35}},\ \bibinfo {pages} {21} (\bibinfo {year} {1991})}\BibitemShut
  {NoStop}%
\bibitem [{\citenamefont {Riley}\ \emph {et~al.}(2019)\citenamefont {Riley}
  \emph {et~al.}}]{Riley:2019yda}%
  \BibitemOpen
  \bibfield  {author} {\bibinfo {author} {\bibfnamefont {T.~E.}\ \bibnamefont
  {Riley}} \emph {et~al.},\ }\href {\doibase 10.3847/2041-8213/ab481c}
  {\bibfield  {journal} {\bibinfo  {journal} {Astrophys. J. Lett.}\ }\textbf
  {\bibinfo {volume} {887}},\ \bibinfo {pages} {L21} (\bibinfo {year}
  {2019})},\ \Eprint {http://arxiv.org/abs/1912.05702} {arXiv:1912.05702
  [astro-ph.HE]} \BibitemShut {NoStop}%
\bibitem [{\citenamefont {Miller}\ \emph {et~al.}(2019)\citenamefont {Miller}
  \emph {et~al.}}]{Miller:2019cac}%
  \BibitemOpen
  \bibfield  {author} {\bibinfo {author} {\bibfnamefont {M.~C.}\ \bibnamefont
  {Miller}} \emph {et~al.},\ }\href {\doibase 10.3847/2041-8213/ab50c5}
  {\bibfield  {journal} {\bibinfo  {journal} {Astrophys. J. Lett.}\ }\textbf
  {\bibinfo {volume} {887}},\ \bibinfo {pages} {L24} (\bibinfo {year}
  {2019})},\ \Eprint {http://arxiv.org/abs/1912.05705} {arXiv:1912.05705
  [astro-ph.HE]} \BibitemShut {NoStop}%
\bibitem [{\citenamefont {Miller}\ \emph {et~al.}(2021)\citenamefont {Miller}
  \emph {et~al.}}]{Miller:2021qha}%
  \BibitemOpen
  \bibfield  {author} {\bibinfo {author} {\bibfnamefont {M.~C.}\ \bibnamefont
  {Miller}} \emph {et~al.},\ }\href {\doibase 10.3847/2041-8213/ac089b}
  {\bibfield  {journal} {\bibinfo  {journal} {Astrophys. J. Lett.}\ }\textbf
  {\bibinfo {volume} {918}},\ \bibinfo {pages} {L28} (\bibinfo {year}
  {2021})},\ \Eprint {http://arxiv.org/abs/2105.06979} {arXiv:2105.06979
  [astro-ph.HE]} \BibitemShut {NoStop}%
\bibitem [{\citenamefont {Vinciguerra}\ \emph {et~al.}(2024)\citenamefont
  {Vinciguerra} \emph {et~al.}}]{Vinciguerra:2023qxq}%
  \BibitemOpen
  \bibfield  {author} {\bibinfo {author} {\bibfnamefont {S.}~\bibnamefont
  {Vinciguerra}} \emph {et~al.},\ }\href {\doibase 10.3847/1538-4357/acfb83}
  {\bibfield  {journal} {\bibinfo  {journal} {Astrophys. J.}\ }\textbf
  {\bibinfo {volume} {961}},\ \bibinfo {pages} {62} (\bibinfo {year} {2024})},\
  \Eprint {http://arxiv.org/abs/2308.09469} {arXiv:2308.09469 [astro-ph.HE]}
  \BibitemShut {NoStop}%
\bibitem [{\citenamefont {Raaijmakers}\ \emph {et~al.}(2021)\citenamefont
  {Raaijmakers}, \citenamefont {Greif}, \citenamefont {Hebeler}, \citenamefont
  {Hinderer}, \citenamefont {Nissanke}, \citenamefont {Schwenk}, \citenamefont
  {Riley}, \citenamefont {Watts}, \citenamefont {Lattimer},\ and\ \citenamefont
  {Ho}}]{Raaijmakers:2021uju}%
  \BibitemOpen
  \bibfield  {author} {\bibinfo {author} {\bibfnamefont {G.}~\bibnamefont
  {Raaijmakers}}, \bibinfo {author} {\bibfnamefont {S.~K.}\ \bibnamefont
  {Greif}}, \bibinfo {author} {\bibfnamefont {K.}~\bibnamefont {Hebeler}},
  \bibinfo {author} {\bibfnamefont {T.}~\bibnamefont {Hinderer}}, \bibinfo
  {author} {\bibfnamefont {S.}~\bibnamefont {Nissanke}}, \bibinfo {author}
  {\bibfnamefont {A.}~\bibnamefont {Schwenk}}, \bibinfo {author} {\bibfnamefont
  {T.~E.}\ \bibnamefont {Riley}}, \bibinfo {author} {\bibfnamefont {A.~L.}\
  \bibnamefont {Watts}}, \bibinfo {author} {\bibfnamefont {J.~M.}\ \bibnamefont
  {Lattimer}}, \ and\ \bibinfo {author} {\bibfnamefont {W.~C.~G.}\ \bibnamefont
  {Ho}},\ }\href {\doibase 10.3847/2041-8213/ac089a} {\bibfield  {journal}
  {\bibinfo  {journal} {Astrophys. J. Lett.}\ }\textbf {\bibinfo {volume}
  {918}},\ \bibinfo {pages} {L29} (\bibinfo {year} {2021})},\ \Eprint
  {http://arxiv.org/abs/2105.06981} {arXiv:2105.06981 [astro-ph.HE]}
  \BibitemShut {NoStop}%
\bibitem [{\citenamefont {Hu}\ \emph {et~al.}(2021)\citenamefont {Hu},
  \citenamefont {Gao}, \citenamefont {Xu},\ and\ \citenamefont
  {Shao}}]{Hu:2021tyw}%
  \BibitemOpen
  \bibfield  {author} {\bibinfo {author} {\bibfnamefont {Z.}~\bibnamefont
  {Hu}}, \bibinfo {author} {\bibfnamefont {Y.}~\bibnamefont {Gao}}, \bibinfo
  {author} {\bibfnamefont {R.}~\bibnamefont {Xu}}, \ and\ \bibinfo {author}
  {\bibfnamefont {L.}~\bibnamefont {Shao}},\ }\href {\doibase
  10.1103/PhysRevD.104.104014} {\bibfield  {journal} {\bibinfo  {journal}
  {Phys. Rev. D}\ }\textbf {\bibinfo {volume} {104}},\ \bibinfo {pages}
  {104014} (\bibinfo {year} {2021})},\ \Eprint
  {http://arxiv.org/abs/2109.13453} {arXiv:2109.13453 [gr-qc]} \BibitemShut
  {NoStop}%
\bibitem [{\citenamefont {Kennedy}\ \emph {et~al.}(2022)\citenamefont {Kennedy}
  \emph {et~al.}}]{Kennedy:2022zml}%
  \BibitemOpen
  \bibfield  {author} {\bibinfo {author} {\bibfnamefont {M.~R.}\ \bibnamefont
  {Kennedy}} \emph {et~al.},\ }\href {\doibase 10.1093/mnras/stac379}
  {\bibfield  {journal} {\bibinfo  {journal} {Mon. Not. Roy. Astron. Soc.}\
  }\textbf {\bibinfo {volume} {512}},\ \bibinfo {pages} {3001} (\bibinfo {year}
  {2022})},\ \Eprint {http://arxiv.org/abs/2202.05111} {arXiv:2202.05111
  [astro-ph.HE]} \BibitemShut {NoStop}%
\bibitem [{\citenamefont {Miao}\ \emph {et~al.}(2022)\citenamefont {Miao},
  \citenamefont {Zhu}, \citenamefont {Li},\ and\ \citenamefont
  {Huang}}]{Miao:2022rqj}%
  \BibitemOpen
  \bibfield  {author} {\bibinfo {author} {\bibfnamefont {Z.}~\bibnamefont
  {Miao}}, \bibinfo {author} {\bibfnamefont {Y.}~\bibnamefont {Zhu}}, \bibinfo
  {author} {\bibfnamefont {A.}~\bibnamefont {Li}}, \ and\ \bibinfo {author}
  {\bibfnamefont {F.}~\bibnamefont {Huang}},\ }\href {\doibase
  10.3847/1538-4357/ac8544} {\bibfield  {journal} {\bibinfo  {journal}
  {Astrophys. J.}\ }\textbf {\bibinfo {volume} {936}},\ \bibinfo {pages} {69}
  (\bibinfo {year} {2022})},\ \Eprint {http://arxiv.org/abs/2204.05560}
  {arXiv:2204.05560 [astro-ph.HE]} \BibitemShut {NoStop}%
\bibitem [{\citenamefont {Shawqi}\ and\ \citenamefont
  {Morsink}(2024)}]{shawqi2024interpreting}%
  \BibitemOpen
  \bibfield  {author} {\bibinfo {author} {\bibfnamefont {S.}~\bibnamefont
  {Shawqi}}\ and\ \bibinfo {author} {\bibfnamefont {S.~M.}\ \bibnamefont
  {Morsink}},\ }\href@noop {} {\enquote {\bibinfo {title} {Interpreting mass
  and radius measurements of neutron stars with dark matter halos},}\ }
  (\bibinfo {year} {2024}),\ \Eprint {http://arxiv.org/abs/2406.03332}
  {arXiv:2406.03332 [astro-ph.HE]} \BibitemShut {NoStop}%
\bibitem [{\citenamefont {Rutledge}\ \emph {et~al.}(2001)\citenamefont
  {Rutledge}, \citenamefont {Bildsten}, \citenamefont {Brown}, \citenamefont
  {Pavlov},\ and\ \citenamefont {Zavlin}}]{Rutledge:2001kc}%
  \BibitemOpen
  \bibfield  {author} {\bibinfo {author} {\bibfnamefont {R.~E.}\ \bibnamefont
  {Rutledge}}, \bibinfo {author} {\bibfnamefont {L.}~\bibnamefont {Bildsten}},
  \bibinfo {author} {\bibfnamefont {E.~F.}\ \bibnamefont {Brown}}, \bibinfo
  {author} {\bibfnamefont {G.~G.}\ \bibnamefont {Pavlov}}, \ and\ \bibinfo
  {author} {\bibfnamefont {V.~E.}\ \bibnamefont {Zavlin}},\ }\href {\doibase
  10.1086/322361} {\bibfield  {journal} {\bibinfo  {journal} {Astrophys. J.}\
  }\textbf {\bibinfo {volume} {559}},\ \bibinfo {pages} {1054} (\bibinfo {year}
  {2001})},\ \Eprint {http://arxiv.org/abs/astro-ph/0105319}
  {arXiv:astro-ph/0105319} \BibitemShut {NoStop}%
\bibitem [{\citenamefont {Tolman}(1939)}]{Tolman:1939jz}%
  \BibitemOpen
  \bibfield  {author} {\bibinfo {author} {\bibfnamefont {R.~C.}\ \bibnamefont
  {Tolman}},\ }\href {\doibase 10.1103/PhysRev.55.364} {\bibfield  {journal}
  {\bibinfo  {journal} {Phys. Rev.}\ }\textbf {\bibinfo {volume} {55}},\
  \bibinfo {pages} {364} (\bibinfo {year} {1939})}\BibitemShut {NoStop}%
\bibitem [{\citenamefont {Oppenheimer}\ and\ \citenamefont
  {Volkoff}(1939)}]{Oppenheimer:1939ne}%
  \BibitemOpen
  \bibfield  {author} {\bibinfo {author} {\bibfnamefont {J.~R.}\ \bibnamefont
  {Oppenheimer}}\ and\ \bibinfo {author} {\bibfnamefont {G.~M.}\ \bibnamefont
  {Volkoff}},\ }\href {\doibase 10.1103/PhysRev.55.374} {\bibfield  {journal}
  {\bibinfo  {journal} {Phys. Rev.}\ }\textbf {\bibinfo {volume} {55}},\
  \bibinfo {pages} {374} (\bibinfo {year} {1939})}\BibitemShut {NoStop}%
\bibitem [{\citenamefont {Sandin}\ and\ \citenamefont
  {Ciarcelluti}(2009)}]{Sandin:2008db}%
  \BibitemOpen
  \bibfield  {author} {\bibinfo {author} {\bibfnamefont {F.}~\bibnamefont
  {Sandin}}\ and\ \bibinfo {author} {\bibfnamefont {P.}~\bibnamefont
  {Ciarcelluti}},\ }\href {\doibase 10.1016/j.astropartphys.2009.09.005}
  {\bibfield  {journal} {\bibinfo  {journal} {Astropart. Phys.}\ }\textbf
  {\bibinfo {volume} {32}},\ \bibinfo {pages} {278} (\bibinfo {year} {2009})},\
  \Eprint {http://arxiv.org/abs/0809.2942} {arXiv:0809.2942 [astro-ph]}
  \BibitemShut {NoStop}%
\bibitem [{\citenamefont {Munoz}(2004)}]{Munoz:2003gx}%
  \BibitemOpen
  \bibfield  {author} {\bibinfo {author} {\bibfnamefont {C.}~\bibnamefont
  {Munoz}},\ }\href {\doibase 10.1142/S0217751X04018154} {\bibfield  {journal}
  {\bibinfo  {journal} {Int. J. Mod. Phys. A}\ }\textbf {\bibinfo {volume}
  {19}},\ \bibinfo {pages} {3093} (\bibinfo {year} {2004})},\ \Eprint
  {http://arxiv.org/abs/hep-ph/0309346} {arXiv:hep-ph/0309346} \BibitemShut
  {NoStop}%
\bibitem [{\citenamefont {Shapiro}\ and\ \citenamefont
  {Teukolsky}(1983)}]{Shapiro:1983du}%
  \BibitemOpen
  \bibfield  {author} {\bibinfo {author} {\bibfnamefont {S.~L.}\ \bibnamefont
  {Shapiro}}\ and\ \bibinfo {author} {\bibfnamefont {S.~A.}\ \bibnamefont
  {Teukolsky}},\ }\href {\doibase 10.1002/9783527617661} {\emph {\bibinfo
  {title} {{Black Holes, White Dwarfs, and Neutron Stars: The Physics of
  Compact Objects}}}}\ (\bibinfo {year} {1983})\BibitemShut {NoStop}%
\bibitem [{\citenamefont {Akmal}\ and\ \citenamefont
  {Pandharipande}(1997)}]{Akmal:1997ft}%
  \BibitemOpen
  \bibfield  {author} {\bibinfo {author} {\bibfnamefont {A.}~\bibnamefont
  {Akmal}}\ and\ \bibinfo {author} {\bibfnamefont {V.~R.}\ \bibnamefont
  {Pandharipande}},\ }\href {\doibase 10.1103/PhysRevC.56.2261} {\bibfield
  {journal} {\bibinfo  {journal} {Phys. Rev. C}\ }\textbf {\bibinfo {volume}
  {56}},\ \bibinfo {pages} {2261} (\bibinfo {year} {1997})},\ \Eprint
  {http://arxiv.org/abs/nucl-th/9705013} {arXiv:nucl-th/9705013} \BibitemShut
  {NoStop}%
\bibitem [{\citenamefont {Li}\ \emph {et~al.}(2022)\citenamefont {Li},
  \citenamefont {Gao}, \citenamefont {Shao}, \citenamefont {Xu},\ and\
  \citenamefont {Xu}}]{Li:2022qql}%
  \BibitemOpen
  \bibfield  {author} {\bibinfo {author} {\bibfnamefont {H.-B.}\ \bibnamefont
  {Li}}, \bibinfo {author} {\bibfnamefont {Y.}~\bibnamefont {Gao}}, \bibinfo
  {author} {\bibfnamefont {L.}~\bibnamefont {Shao}}, \bibinfo {author}
  {\bibfnamefont {R.-X.}\ \bibnamefont {Xu}}, \ and\ \bibinfo {author}
  {\bibfnamefont {R.}~\bibnamefont {Xu}},\ }\href {\doibase
  10.1093/mnras/stac2622} {\bibfield  {journal} {\bibinfo  {journal} {Mon. Not.
  Roy. Astron. Soc.}\ }\textbf {\bibinfo {volume} {516}},\ \bibinfo {pages}
  {6172} (\bibinfo {year} {2022})},\ \Eprint {http://arxiv.org/abs/2206.09407}
  {arXiv:2206.09407 [gr-qc]} \BibitemShut {NoStop}%
\bibitem [{\citenamefont {Henriques}\ \emph {et~al.}(1990)\citenamefont
  {Henriques}, \citenamefont {Liddle},\ and\ \citenamefont
  {Moorhouse}}]{Henriques:1990xg}%
  \BibitemOpen
  \bibfield  {author} {\bibinfo {author} {\bibfnamefont {A.~B.}\ \bibnamefont
  {Henriques}}, \bibinfo {author} {\bibfnamefont {A.~R.}\ \bibnamefont
  {Liddle}}, \ and\ \bibinfo {author} {\bibfnamefont {R.~G.}\ \bibnamefont
  {Moorhouse}},\ }\href {\doibase 10.1016/0370-2693(90)90789-9} {\bibfield
  {journal} {\bibinfo  {journal} {Phys. Lett. B}\ }\textbf {\bibinfo {volume}
  {251}},\ \bibinfo {pages} {511} (\bibinfo {year} {1990})}\BibitemShut
  {NoStop}%
\bibitem [{\citenamefont {Jetzer}(1990)}]{Jetzer:1990xa}%
  \BibitemOpen
  \bibfield  {author} {\bibinfo {author} {\bibfnamefont {P.}~\bibnamefont
  {Jetzer}},\ }\href {\doibase 10.1016/0370-2693(90)90952-3} {\bibfield
  {journal} {\bibinfo  {journal} {Phys. Lett. B}\ }\textbf {\bibinfo {volume}
  {243}},\ \bibinfo {pages} {36} (\bibinfo {year} {1990})}\BibitemShut
  {NoStop}%
\bibitem [{\citenamefont {Riley}\ \emph {et~al.}(2021)\citenamefont {Riley}
  \emph {et~al.}}]{Riley:2021pdl}%
  \BibitemOpen
  \bibfield  {author} {\bibinfo {author} {\bibfnamefont {T.~E.}\ \bibnamefont
  {Riley}} \emph {et~al.},\ }\href {\doibase 10.3847/2041-8213/ac0a81}
  {\bibfield  {journal} {\bibinfo  {journal} {Astrophys. J. Lett.}\ }\textbf
  {\bibinfo {volume} {918}},\ \bibinfo {pages} {L27} (\bibinfo {year}
  {2021})},\ \Eprint {http://arxiv.org/abs/2105.06980} {arXiv:2105.06980
  [astro-ph.HE]} \BibitemShut {NoStop}%
\bibitem [{\citenamefont {Salmi}\ \emph {et~al.}(2024)\citenamefont {Salmi}
  \emph {et~al.}}]{Salmi:2024aum}%
  \BibitemOpen
  \bibfield  {author} {\bibinfo {author} {\bibfnamefont {T.}~\bibnamefont
  {Salmi}} \emph {et~al.},\ }\href@noop {} {\  (\bibinfo {year} {2024})},\
  \Eprint {http://arxiv.org/abs/2406.14466} {arXiv:2406.14466 [astro-ph.HE]}
  \BibitemShut {NoStop}%
\bibitem [{\citenamefont {Amaro-Seoane}\ \emph {et~al.}(2010)\citenamefont
  {Amaro-Seoane}, \citenamefont {Barranco}, \citenamefont {Bernal},\ and\
  \citenamefont {Rezzolla}}]{Amaro-Seoane:2010pks}%
  \BibitemOpen
  \bibfield  {author} {\bibinfo {author} {\bibfnamefont {P.}~\bibnamefont
  {Amaro-Seoane}}, \bibinfo {author} {\bibfnamefont {J.}~\bibnamefont
  {Barranco}}, \bibinfo {author} {\bibfnamefont {A.}~\bibnamefont {Bernal}}, \
  and\ \bibinfo {author} {\bibfnamefont {L.}~\bibnamefont {Rezzolla}},\ }\href
  {\doibase 10.1088/1475-7516/2010/11/002} {\bibfield  {journal} {\bibinfo
  {journal} {JCAP}\ }\textbf {\bibinfo {volume} {11}},\ \bibinfo {pages} {002}
  (\bibinfo {year} {2010})},\ \Eprint {http://arxiv.org/abs/1009.0019}
  {arXiv:1009.0019 [astro-ph.CO]} \BibitemShut {NoStop}%
\bibitem [{\citenamefont {Choudhury}\ \emph {et~al.}(2024)\citenamefont
  {Choudhury} \emph {et~al.}}]{Choudhury:2024xbk}%
  \BibitemOpen
  \bibfield  {author} {\bibinfo {author} {\bibfnamefont {D.}~\bibnamefont
  {Choudhury}} \emph {et~al.},\ }\href {\doibase 10.3847/2041-8213/ad5a6f}
  {\bibfield  {journal} {\bibinfo  {journal} {Astrophys. J. Lett.}\ }\textbf
  {\bibinfo {volume} {971}},\ \bibinfo {pages} {L20} (\bibinfo {year}
  {2024})},\ \Eprint {http://arxiv.org/abs/2407.06789} {arXiv:2407.06789
  [astro-ph.HE]} \BibitemShut {NoStop}%
\bibitem [{\citenamefont {Doroshenko}\ \emph {et~al.}(2022)\citenamefont
  {Doroshenko}, \citenamefont {Suleimanov}, \citenamefont {P\"uhlhofer},\ and\
  \citenamefont {Santangelo}}]{Doroshenko:2022nwp}%
  \BibitemOpen
  \bibfield  {author} {\bibinfo {author} {\bibfnamefont {V.}~\bibnamefont
  {Doroshenko}}, \bibinfo {author} {\bibfnamefont {V.}~\bibnamefont
  {Suleimanov}}, \bibinfo {author} {\bibfnamefont {G.}~\bibnamefont
  {P\"uhlhofer}}, \ and\ \bibinfo {author} {\bibfnamefont {A.}~\bibnamefont
  {Santangelo}},\ }\href {\doibase 10.1038/s41550-022-01800-1} {\bibfield
  {journal} {\bibinfo  {journal} {Nature Astron.}\ }\textbf {\bibinfo {volume}
  {6}},\ \bibinfo {pages} {1444} (\bibinfo {year} {2022})}\BibitemShut
  {NoStop}%
\bibitem [{\citenamefont {Silva}\ and\ \citenamefont
  {Yunes}(2019)}]{Silva:2018yxz}%
  \BibitemOpen
  \bibfield  {author} {\bibinfo {author} {\bibfnamefont {H.~O.}\ \bibnamefont
  {Silva}}\ and\ \bibinfo {author} {\bibfnamefont {N.}~\bibnamefont {Yunes}},\
  }\href {\doibase 10.1103/PhysRevD.99.044034} {\bibfield  {journal} {\bibinfo
  {journal} {Phys. Rev. D}\ }\textbf {\bibinfo {volume} {99}},\ \bibinfo
  {pages} {044034} (\bibinfo {year} {2019})},\ \Eprint
  {http://arxiv.org/abs/1808.04391} {arXiv:1808.04391 [gr-qc]} \BibitemShut
  {NoStop}%
\bibitem [{\citenamefont {Xu}\ \emph {et~al.}(2020)\citenamefont {Xu},
  \citenamefont {Gao},\ and\ \citenamefont {Shao}}]{Xu:2020vbs}%
  \BibitemOpen
  \bibfield  {author} {\bibinfo {author} {\bibfnamefont {R.}~\bibnamefont
  {Xu}}, \bibinfo {author} {\bibfnamefont {Y.}~\bibnamefont {Gao}}, \ and\
  \bibinfo {author} {\bibfnamefont {L.}~\bibnamefont {Shao}},\ }\href {\doibase
  10.1103/PhysRevD.102.064057} {\bibfield  {journal} {\bibinfo  {journal}
  {Phys. Rev. D}\ }\textbf {\bibinfo {volume} {102}},\ \bibinfo {pages}
  {064057} (\bibinfo {year} {2020})},\ \Eprint
  {http://arxiv.org/abs/2007.10080} {arXiv:2007.10080 [gr-qc]} \BibitemShut
  {NoStop}%
\bibitem [{\citenamefont {Miller}\ \emph {et~al.}(1998)\citenamefont {Miller},
  \citenamefont {Lamb},\ and\ \citenamefont {Cook}}]{Miller:1998gr}%
  \BibitemOpen
  \bibfield  {author} {\bibinfo {author} {\bibfnamefont {M.~C.}\ \bibnamefont
  {Miller}}, \bibinfo {author} {\bibfnamefont {F.~K.}\ \bibnamefont {Lamb}}, \
  and\ \bibinfo {author} {\bibfnamefont {G.~B.}\ \bibnamefont {Cook}},\ }\href
  {\doibase 10.1086/306533} {\bibfield  {journal} {\bibinfo  {journal}
  {Astrophys. J.}\ }\textbf {\bibinfo {volume} {509}},\ \bibinfo {pages} {793}
  (\bibinfo {year} {1998})},\ \Eprint {http://arxiv.org/abs/astro-ph/9805007}
  {arXiv:astro-ph/9805007} \BibitemShut {NoStop}%
\bibitem [{\citenamefont {Morsink}\ \emph {et~al.}(2007)\citenamefont
  {Morsink}, \citenamefont {Leahy}, \citenamefont {Cadeau},\ and\ \citenamefont
  {Braga}}]{Morsink:2007tv}%
  \BibitemOpen
  \bibfield  {author} {\bibinfo {author} {\bibfnamefont {S.~M.}\ \bibnamefont
  {Morsink}}, \bibinfo {author} {\bibfnamefont {D.~A.}\ \bibnamefont {Leahy}},
  \bibinfo {author} {\bibfnamefont {C.}~\bibnamefont {Cadeau}}, \ and\ \bibinfo
  {author} {\bibfnamefont {J.}~\bibnamefont {Braga}},\ }\href {\doibase
  10.1086/518648} {\bibfield  {journal} {\bibinfo  {journal} {Astrophys. J.}\
  }\textbf {\bibinfo {volume} {663}},\ \bibinfo {pages} {1244} (\bibinfo {year}
  {2007})},\ \Eprint {http://arxiv.org/abs/astro-ph/0703123}
  {arXiv:astro-ph/0703123} \BibitemShut {NoStop}%
\bibitem [{\citenamefont {Poutanen}\ and\ \citenamefont
  {Beloborodov}(2006)}]{Poutanen:2006hw}%
  \BibitemOpen
  \bibfield  {author} {\bibinfo {author} {\bibfnamefont {J.}~\bibnamefont
  {Poutanen}}\ and\ \bibinfo {author} {\bibfnamefont {A.~M.}\ \bibnamefont
  {Beloborodov}},\ }\href {\doibase 10.1111/j.1365-2966.2006.11088.x}
  {\bibfield  {journal} {\bibinfo  {journal} {Mon. Not. Roy. Astron. Soc.}\
  }\textbf {\bibinfo {volume} {373}},\ \bibinfo {pages} {836} (\bibinfo {year}
  {2006})},\ \Eprint {http://arxiv.org/abs/astro-ph/0608663}
  {arXiv:astro-ph/0608663} \BibitemShut {NoStop}%
\bibitem [{\citenamefont {Rybicki}(2004)}]{Rybicki:2004hfl}%
  \BibitemOpen
  \bibfield  {author} {\bibinfo {author} {\bibfnamefont {G.~B.}\ \bibnamefont
  {Rybicki}},\ }\href {\doibase 10.1002/9783527618170} {\emph {\bibinfo {title}
  {{Radiative Processes in Astrophysics}}}}\ (\bibinfo  {publisher}
  {Wiley-VCH},\ \bibinfo {year} {2004})\BibitemShut {NoStop}%
\bibitem [{\citenamefont {{Pechenick}}\ \emph {et~al.}(1983)\citenamefont
  {{Pechenick}}, \citenamefont {{Ftaclas}},\ and\ \citenamefont
  {{Cohen}}}]{1983ApJ...274..846P}%
  \BibitemOpen
  \bibfield  {author} {\bibinfo {author} {\bibfnamefont {K.~R.}\ \bibnamefont
  {{Pechenick}}}, \bibinfo {author} {\bibfnamefont {C.}~\bibnamefont
  {{Ftaclas}}}, \ and\ \bibinfo {author} {\bibfnamefont {J.~M.}\ \bibnamefont
  {{Cohen}}},\ }\href {\doibase 10.1086/161498} {\bibfield  {journal} {\bibinfo
   {journal} {\apj}\ }\textbf {\bibinfo {volume} {274}},\ \bibinfo {pages}
  {846} (\bibinfo {year} {1983})}\BibitemShut {NoStop}%
\bibitem [{\citenamefont {Bogdanov}\ \emph {et~al.}(2008)\citenamefont
  {Bogdanov}, \citenamefont {Grindlay},\ and\ \citenamefont
  {Rybicki}}]{Bogdanov:2008qm}%
  \BibitemOpen
  \bibfield  {author} {\bibinfo {author} {\bibfnamefont {S.}~\bibnamefont
  {Bogdanov}}, \bibinfo {author} {\bibfnamefont {J.~E.}\ \bibnamefont
  {Grindlay}}, \ and\ \bibinfo {author} {\bibfnamefont {G.~B.}\ \bibnamefont
  {Rybicki}},\ }\href {\doibase 10.1086/592341} {\bibfield  {journal} {\bibinfo
   {journal} {Astrophys. J.}\ }\textbf {\bibinfo {volume} {689}},\ \bibinfo
  {pages} {407} (\bibinfo {year} {2008})},\ \Eprint
  {http://arxiv.org/abs/0801.4030} {arXiv:0801.4030 [astro-ph]} \BibitemShut
  {NoStop}%
\bibitem [{\citenamefont {Abbott}\ \emph {et~al.}(2017)\citenamefont {Abbott}
  \emph {et~al.}}]{LIGOScientific:2017vwq}%
  \BibitemOpen
  \bibfield  {author} {\bibinfo {author} {\bibfnamefont {B.~P.}\ \bibnamefont
  {Abbott}} \emph {et~al.} (\bibinfo {collaboration} {LIGO Scientific,
  Virgo}),\ }\href {\doibase 10.1103/PhysRevLett.119.161101} {\bibfield
  {journal} {\bibinfo  {journal} {Phys. Rev. Lett.}\ }\textbf {\bibinfo
  {volume} {119}},\ \bibinfo {pages} {161101} (\bibinfo {year} {2017})},\
  \Eprint {http://arxiv.org/abs/1710.05832} {arXiv:1710.05832 [gr-qc]}
  \BibitemShut {NoStop}%
\bibitem [{\citenamefont {Bogdanov}\ \emph {et~al.}(2019)\citenamefont
  {Bogdanov} \emph {et~al.}}]{Bogdanov:2019qjb}%
  \BibitemOpen
  \bibfield  {author} {\bibinfo {author} {\bibfnamefont {S.}~\bibnamefont
  {Bogdanov}} \emph {et~al.},\ }\href {\doibase 10.3847/2041-8213/ab5968}
  {\bibfield  {journal} {\bibinfo  {journal} {Astrophys. J. Lett.}\ }\textbf
  {\bibinfo {volume} {887}},\ \bibinfo {pages} {L26} (\bibinfo {year}
  {2019})},\ \Eprint {http://arxiv.org/abs/1912.05707} {arXiv:1912.05707
  [astro-ph.HE]} \BibitemShut {NoStop}%
\bibitem [{\citenamefont {Wolff}\ \emph {et~al.}(2021)\citenamefont {Wolff}
  \emph {et~al.}}]{Wolff:2021oba}%
  \BibitemOpen
  \bibfield  {author} {\bibinfo {author} {\bibfnamefont {M.~T.}\ \bibnamefont
  {Wolff}} \emph {et~al.},\ }\href {\doibase 10.3847/2041-8213/ac158e}
  {\bibfield  {journal} {\bibinfo  {journal} {Astrophys. J. Lett.}\ }\textbf
  {\bibinfo {volume} {918}},\ \bibinfo {pages} {L26} (\bibinfo {year}
  {2021})},\ \Eprint {http://arxiv.org/abs/2105.06978} {arXiv:2105.06978
  [astro-ph.HE]} \BibitemShut {NoStop}%
\bibitem [{\citenamefont {Li}\ \emph {et~al.}(2012)\citenamefont {Li},
  \citenamefont {Wang},\ and\ \citenamefont {Cheng}}]{Li:2012qf}%
  \BibitemOpen
  \bibfield  {author} {\bibinfo {author} {\bibfnamefont {X.}~\bibnamefont
  {Li}}, \bibinfo {author} {\bibfnamefont {F.}~\bibnamefont {Wang}}, \ and\
  \bibinfo {author} {\bibfnamefont {K.~S.}\ \bibnamefont {Cheng}},\ }\href
  {\doibase 10.1088/1475-7516/2012/10/031} {\bibfield  {journal} {\bibinfo
  {journal} {JCAP}\ }\textbf {\bibinfo {volume} {10}},\ \bibinfo {pages} {031}
  (\bibinfo {year} {2012})},\ \Eprint {http://arxiv.org/abs/1210.1748}
  {arXiv:1210.1748 [astro-ph.CO]} \BibitemShut {NoStop}%
\bibitem [{\citenamefont {Boehmer}\ and\ \citenamefont
  {Harko}(2007)}]{Boehmer:2007um}%
  \BibitemOpen
  \bibfield  {author} {\bibinfo {author} {\bibfnamefont {C.~G.}\ \bibnamefont
  {Boehmer}}\ and\ \bibinfo {author} {\bibfnamefont {T.}~\bibnamefont
  {Harko}},\ }\href {\doibase 10.1088/1475-7516/2007/06/025} {\bibfield
  {journal} {\bibinfo  {journal} {JCAP}\ }\textbf {\bibinfo {volume} {06}},\
  \bibinfo {pages} {025} (\bibinfo {year} {2007})},\ \Eprint
  {http://arxiv.org/abs/0705.4158} {arXiv:0705.4158 [astro-ph]} \BibitemShut
  {NoStop}%
\bibitem [{\citenamefont {Mariani}\ \emph {et~al.}(2024)\citenamefont
  {Mariani}, \citenamefont {Albertus}, \citenamefont {del
  Rosario~Alessandroni}, \citenamefont {Orsaria}, \citenamefont
  {Perez-Garcia},\ and\ \citenamefont {Ranea-Sandoval}}]{Mariani:2023wtv}%
  \BibitemOpen
  \bibfield  {author} {\bibinfo {author} {\bibfnamefont {M.}~\bibnamefont
  {Mariani}}, \bibinfo {author} {\bibfnamefont {C.}~\bibnamefont {Albertus}},
  \bibinfo {author} {\bibfnamefont {M.}~\bibnamefont {del
  Rosario~Alessandroni}}, \bibinfo {author} {\bibfnamefont {M.~G.}\
  \bibnamefont {Orsaria}}, \bibinfo {author} {\bibfnamefont {M.~A.}\
  \bibnamefont {Perez-Garcia}}, \ and\ \bibinfo {author} {\bibfnamefont
  {I.~F.}\ \bibnamefont {Ranea-Sandoval}},\ }\href {\doibase
  10.1093/mnras/stad3658} {\bibfield  {journal} {\bibinfo  {journal} {Mon. Not.
  Roy. Astron. Soc.}\ }\textbf {\bibinfo {volume} {527}},\ \bibinfo {pages}
  {6795} (\bibinfo {year} {2024})},\ \Eprint {http://arxiv.org/abs/2311.14004}
  {arXiv:2311.14004 [astro-ph.HE]} \BibitemShut {NoStop}%
\end{thebibliography}%

\end{document}